\documentclass[12pt]{article}
\usepackage{amsmath}
\usepackage{graphicx}
\usepackage{url} 

\newcommand{\blind}{0}

\allowdisplaybreaks
\usepackage[round]{natbib}
\usepackage{thmtools, hyperref}
\hypersetup{
colorlinks,
linkcolor = blue,
urlcolor = blue,
citecolor = black
}
\usepackage[nameinlink]{cleveref}
\usepackage{amsfonts}

\usepackage{mathrsfs}
\usepackage{amssymb}
\usepackage{amsthm}
\usepackage{latexsym,amsmath}
\usepackage{graphicx}
\usepackage{afterpage}
\usepackage{enumitem}
\usepackage{subcaption}
\usepackage{lipsum}
\usepackage{tabularray}
\usepackage{booktabs}
\usepackage{tabularx}
\usepackage{adjustbox}
\usepackage{multirow}
\usepackage[title]{appendix}
\usepackage{longtable}
\usepackage{algorithm}
\usepackage{algpseudocode}
\usepackage{dsfont}
\newcommand{\cc}[1]{\mathcal{#1}}

\newcommand{\bb}[1]{\mathbb{#1}}
\newcommand{\bs}[1]{\boldsymbol{#1}}
\newcommand{\tht}{\theta}
\newcommand{\Tht}{\Theta}
\newcommand{\hth}{\hat{\theta}}
\newcommand{\lm}{\lambda}

\newcommand{\logit}{\mathrm{logit}}

\newcommand{\ind}{\mathds{1}}
\newcommand{\thh}{\textsuperscript{th }}

\newcommand{\dd}{\mathrm{d}}

\DeclareMathOperator*{\eqas}{\overset{a.s.}{=}}

\DeclareMathOperator*{\siid}{\overset{iid}{\sim}}
\newcommand{\appropto}{\mathrel{\vcenter{
  \offinterlineskip\halign{\hfil$##$\cr
    \propto\cr\noalign{\kern2pt}\sim\cr\noalign{\kern-2pt}}}}}
\usepackage{accents}

\newcommand{\kl}{\mathrm{KL}}

\usepackage{subfiles}

\usepackage{authblk}

\begin{document}

\def\spacingset#1{\renewcommand{\baselinestretch}%
{#1}\small\normalsize} \spacingset{1}

\crefname{subsection}{subsection}{subsections}
\Crefname{subsection}{Subsection}{Subsections}

\newtheorem{theorem}{Theorem}[section]
\newtheorem{lemma}[theorem]{Lemma}
\newtheorem{proposition}[theorem]{Proposition}
\newtheorem{corollary}[theorem]{Corollary}
\newtheorem{conjecture}[theorem]{Conjecture}
\newtheorem{definition}[theorem]{Definition}
\newtheorem{example}[theorem]{Example}
\newtheorem{remark}[theorem]{Remark}
\newtheorem{question}[theorem]{Question}
\newtheorem{notation}[theorem]{Notation}
\newtheorem{conditions}[theorem]{Conditions}
\numberwithin{equation}{section}
\newtheorem{assumption}[theorem]{Assumption}


\if0\blind
{
\title{\bf Neural Networks for Parameter Estimation of the Discretely Observed Hawkes Process}
\author[1, 2, 3]{Jason J. Lambe}
\author[1]{Feng Chen}
\author[1]{Tom Stindl}
\author[1]{Tsz-Kit Jeffrey Kwan}

\affil[1]{\footnotesize School of Mathematics and Statistics, UNSW Sydney, Australia}
\affil[2]{\footnotesize Defence Science and Technology Group, Sydney, Australia}
\affil[3]{\footnotesize Corresponding Author: \href{mailto:j.lambe@unsw.edu.au}{j.lambe@unsw.edu.au}}
  %
  \maketitle
} \fi

\if1\blind
{
  \bigskip
  \bigskip
  \bigskip
  \begin{center}
    {\LARGE\bf Title}
\end{center}
  \medskip
} \fi

\bigskip
\begin{abstract}
When the sample path of a Hawkes process is observed discretely, such that only the total event counts in disjoint time intervals are known, the likelihood function becomes intractable. To overcome the challenge of likelihood-based inference in this setting, we propose to use a likelihood-free approach that uses simulated data to train a fully connected neural network (NN) to estimate the parameters of the Hawkes process from a summary statistic of the count data. A naive imputation estimate of the parameters forms the basis for our summary statistic, which is fast to generate and requires minimal expert knowledge to design. The resulting NN estimator is comparable to the best extant approximate likelihood estimators in terms of mean-squared error but requires significantly less computational time. We implement NN quantile estimation for fast uncertainty quantification. The proposed estimation procedure is applied to weekly count data for two infectious diseases, with a time-varying background rate used to capture seasonal fluctuations in infection risk.
\end{abstract}

\noindent%
\vfill

\newpage

\section{Introduction}
The Hawkes process \citep{hawkesSpectraSelfexcitingMutually1971} is a stochastic point process model that exhibits \textit{self-excitation}, whereby the occurrence of an event triggers a short-term spike in the arrival rate of subsequent events. It admits an equivalent mathematical formulation as a \textit{cluster process} \citep{hawkesClusterProcessRepresentation1974}, with events divided into two categories: \textit{immigrants} and \textit{offspring}. An immigrant event arrives according to a background rate function and subsequently produces a random number of offspring, with waiting times to the birth of offspring controlled by an offspring density function. The temporal clustering property of the Hawkes process makes it a popular model for many event sequences, such as earthquakes \citep{ogataStatisticalModelsEarthquake1988a}, financial transactions \citep{ChenHall2013InferenceSEPP,clinetStatisticalInferenceErgodic2017a}, neuronal activity \citep{bonnetNeuronalNetworkInference2022}, and terror attacks \citep{junFlexibleMultivariateSpatiotemporal2024}. When all event times are observed over a fixed time period, the parameters of the Hawkes process can be estimated by Maximum Likelihood (ML) \citep{ogataAsymptoticBehaviourMaximum1978, ozakiMaximumLikelihoodEstimation1979}, or via Expectation Maximisation (EM) \citep{chornoboyMaximumLikelihoodIdentification1988}.

However, cost barriers or measurement imprecision may prevent the continuous observation of a Hawkes process sample path.  In such circumstances, one typically has access only to the total event counts in disjoint time intervals, known as \textit{interval censored} or \textit{aggregated} data. The likelihood function of the Hawkes process relative to an interval censored sample path is analytically intractable, so ML or EM estimation techniques are infeasible. Recently, much attention has been devoted to developing useful methods of inference in this setting. An early work is that of \cite{kirchnerEstimationProcedureHawkes2017}, who establishes an approximation of the Hawkes process using an integer-valued autoregression, from which estimates are obtained. \cite{cheyssonSpectralEstimationHawkes2022} derive a Whittle estimator for the process, which is consistent and asymptotically normal. However, this spectral approach is valid only when the data are aggregated into equally sized intervals, and the Hawkes process has a constant background arrival rate. 
\cite{shlomovichUnivariate2022} propose a modified EM algorithm, where, in the E-step, the authors deterministically build a complete sample path of event times that agrees with the observed count data, by selecting the latent event times to be the mode of a proposal distribution on each observation window. The authors claim that this captures the self-excitation of the Hawkes process within and across censoring intervals. The method is extended to the multivariate setting in \cite{shlomovichMultivariate2022}. The estimation procedure exhibits significant bias in general \citep{chenEstimatingHawkesProcess2025, lambeFittingMultivariateHawkes2025}, and no method for uncertainty quantification is given. \cite{schneiderEstimationSelfexcitingPoint2023} presents an iterative estimation method. Starting with an initial parameter, a sample path is simulated, and events are then added and removed according to one of four proposed algorithms such that the final path matches the observed counts. A new parameter estimate is obtained via MLE or EM, with the process repeated until numerical convergence is achieved. The estimation of standard errors is also not addressed in this work, and in general, each of the proposed algorithms results in some bias, showing similar performance to MCEM in simulation experiments \citep{schneiderEstimationSelfexcitingPoint2023}.


A pseudo-marginal Metropolis--Hastings (PMMH) algorithm is proposed by \cite{chenEstimatingHawkesProcess2025}. The intractable likelihood function is estimated using sequential Monte Carlo (SMC), with the true likelihood replaced by the SMC estimate in an otherwise typical Metropolis-Hastings algorithm \citep{metropolisEquationStateCalculations1953, hastingsMonteCarloSampling1970}. The true likelihood is proportional to the density of the stationary distribution of the PMMH chain \citep{andrieuParticleMarkovChain2010}; hence, the final estimates accurately approximate the true MLE from the discretely observed Hawkes process and exhibit very little empirical bias. Standard error estimates are also automatically available from the PMMH sample. This technique is extended to the multivariate case in \cite{lambeFittingMultivariateHawkes2025}, with improvements to the statistical efficiency of the SMC estimates by adopting a proposal distribution for the latent event times that improves the effective sample size of the set of particles. Though PMMH estimation performs very well, it is highly computationally expensive, particularly for non-Markovian Hawkes processes.

\textit{Simulation based} inference techniques offer computationally efficient alternatives to likelihood based inference for intractable models such as the discretely observed Hawkes process. Approximate Bayesian Computation \citep[ABC;][]{tavareInferringCoalescenceTimes1997, beaumontApproximateBayesianComputation2002, sissonSequentialMonteCarlo2007} enables the estimation of posterior distributions by matching a summary of the observed data with summaries of simulated data according to an appropriate metric. \textit{Indirect inference} makes use of an auxiliary, tractable parametric model, with point estimates obtained via the optimisation of a Monte Carlo approximation to some contrast function, such as the Kullback-Leibler ($\kl$) divergence between the true model and the auxiliary model \citep{gourierouxIndirectInference1993}. In modern applications, indirect inference has been utilised within an ABC context, treating estimates of the auxiliary parameter as highly informative summary statistics for use in ABC algorithms \citep{drovandiApproximateBayesianComputation2011, drovandiBayesianIndirectInference2015}. A detailed review of traditional and modern approaches to simulation-based inference is presented in \cite{cranmerFrontierSimulationbasedInference2020}.

Interest has recently been directed towards the use of neural networks (NNs) for simulation based inference. 
The basic concept is to train a NN via supervised learning on a simulated collection of parameter-observation pairs. The trained model then produces point estimates from observed data. Early examples include \cite{chonLinearNonlinearARMA1997}, \cite{tianARParameterEstimation1997}, and \cite{chonRobustNonlinearARMA1999}, who use NNs to estimate the parameters of various autoregressive moving average models. More recently, \cite{jiangLearningSummaryStatistic2017} use a NN to produce a low-dimensional summary statistic of data, which are then used in traditional ABC. \cite{creelNeuralNetsIndirect2017} uses a NN for indirect inference by training the model on auxiliary parameter estimates, using the final output of the NN either as a standalone point estimate or as a summary statistic in ABC. Increasingly complex inference tasks have since been addressed using neural networks, particularly in relation to spatial and spatio-temporal models \citep{zammit-mangionDeepIntegrodifferenceEquation2020, lenziNeuralNetworksParameter2023, sainsbury-daleLikelihoodFreeParameterEstimation2024}. A major advantage of NN estimation over alternatives is that the often significant cost of training the NN is expended only once at the outset, with subsequent inferences being made near instantaneously. This is referred to as \textit{amortised inference} (see \cite{gershmanAmortizedInferenceProbabilistic2014} for a discussion in relation to human probabilistic inferences), which has been exploited for solving optimisation problems at least as far back as \cite{dayanHelmholtzMachine1995}. \cite{rezendeStochasticBackpropagationApproximate2014} and \cite{gregorDeepAutoRegressiveNetworks2014} provide more recent techniques to this effect. 

The available methods for estimating the interval censored Hawkes process present a trade-off between accuracy and computational time. Though \cite{shlomovichUnivariate2022} and \cite{schneiderEstimationSelfexcitingPoint2023} have carefully designed their respective algorithms for reconstructing the latent event times, both methods are fundamentally biased. On the other hand, the computational time of the PMMH algorithm in \cite{chenEstimatingHawkesProcess2025} is prohibitive in applications where speed is important or when working with many observations. In this work, we propose to use a NN estimator of the model parameters that offers accuracy comparable to PMMH, without the heavy computational burden. A fully connected, feed-forward NN is trained to jointly map summary statistics of the count data to point estimates of the parameter and to point estimates of the upper and lower $\tau$-quantiles of the posterior distribution \citep{fisherDeepLearningMarginal2024, sainsbury-daleNeuralBayesEstimators2025}. Estimation and Bayesian uncertainty quantification are thus fully amortised, in the sense that once trained, the network can be applied to new datasets to provide parameter estimates  and the corresponding $1-2\zeta$ credible intervals via a single forward pass, without retraining or additional model fitting.

For the Markovian Hawkes process, our summary statistic is a single imputation estimate of the parameters. The parameter estimates from a Negative Binomial autoregression (NBAR) are additionally included when working with the non-Markovian Hawkes process, for identification of the offspring density. The single imputation estimate provides a highly informative and low-dimensional summary statistic that only requires simple NNs to produce quality results. Our proposed method can handle unequally sized censoring intervals and time-varying background arrival rates, an advantage over many of the extant approximate likelihood methods. Since the dimension of the summary statistic is independent of the length of the sample path, NN estimation is amortised relative to the length of observation, greatly improving the potential return on training cost. Simulation experiments will demonstrate that the NN estimates perform similarly to PMMH estimates in terms of accuracy, whilst providing significant gains in computational speed and scalability.

The remainder of the article is organised as follows. In Section~\ref{sec:nn_data_method}, we describe the Hawkes process and the likelihood that results from interval censoring. A precise formulation of NN point estimation and quantile estimation are also given. Our choice of summary statistic is detailed in Section~\ref{sec:sumstat}, along with a description of our chosen prior and a general discussion of the advantages of using a summary statistic instead of the complete dataset. Section~\ref{sec:nn_simstudy} assesses our method on various specifications of the Hawkes process through simulation experiments, with comparison made to alternative methods in the literature. In Section~\ref{sec:app_infec_dis}, we demonstrate the efficacy of the NN estimates by replicating the analysis of weekly measles cases across Tokyo (2012 -- 2020) performed by \cite{cheyssonSpectralEstimationHawkes2022} and \cite{chenEstimatingHawkesProcess2025}, obtaining similar results to the latter work. Finally, we model Salmonella cases across New South Wales, Australia (2009 -- 2017) using a time-varying background rate. A discussion of our findings and potential avenues for future work is presented in Section~\ref{sec:NN_disc}. The appendix contains additional simulation results and the computer code implementing the proposed methodology.

\section{Data and Methodology}\label{sec:nn_data_method}
\subsection{The Hawkes process}
\begin{figure}[ht]
\begin{subfigure}{0.49\textwidth}
\centering
    \includegraphics[width = \textwidth]{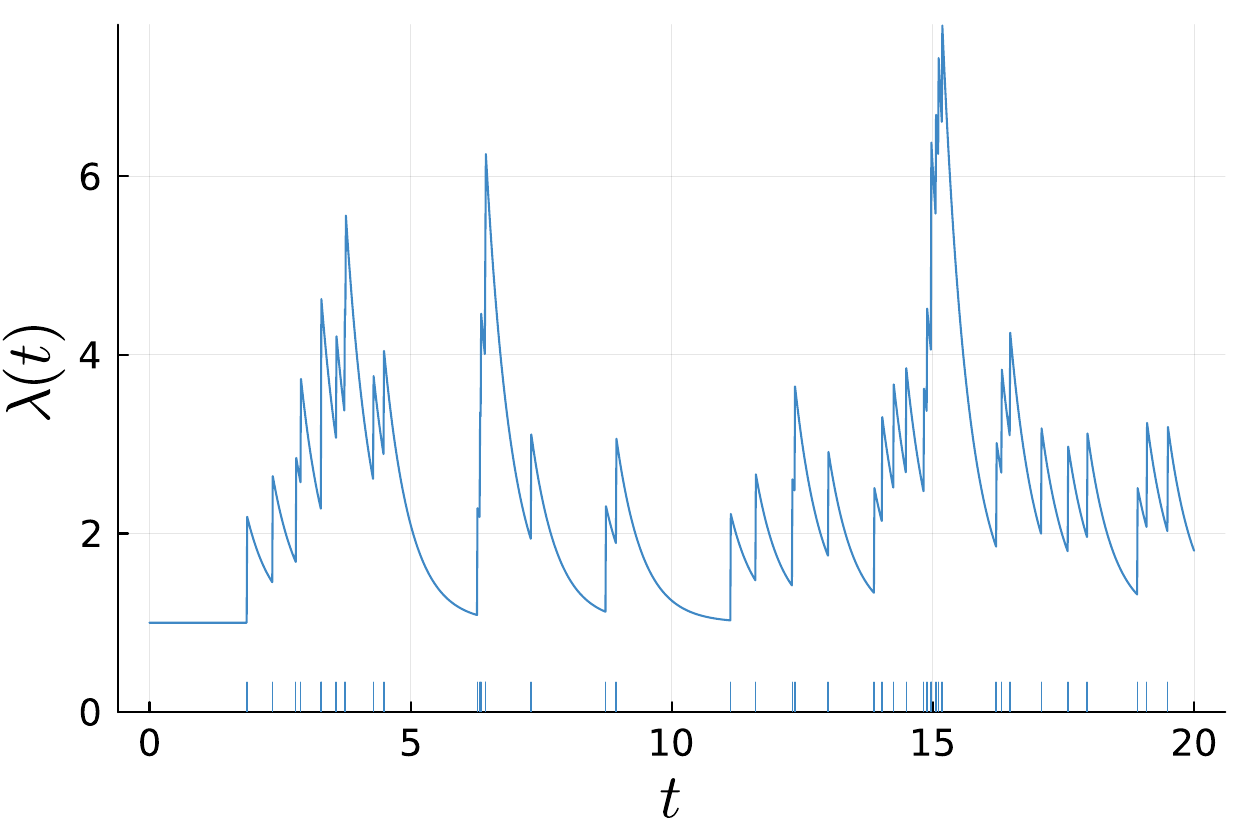}
    \label{fig:lm}
\end{subfigure}
\hfill
\begin{subfigure}{0.49\textwidth}
    \centering
    \includegraphics[width = \textwidth]{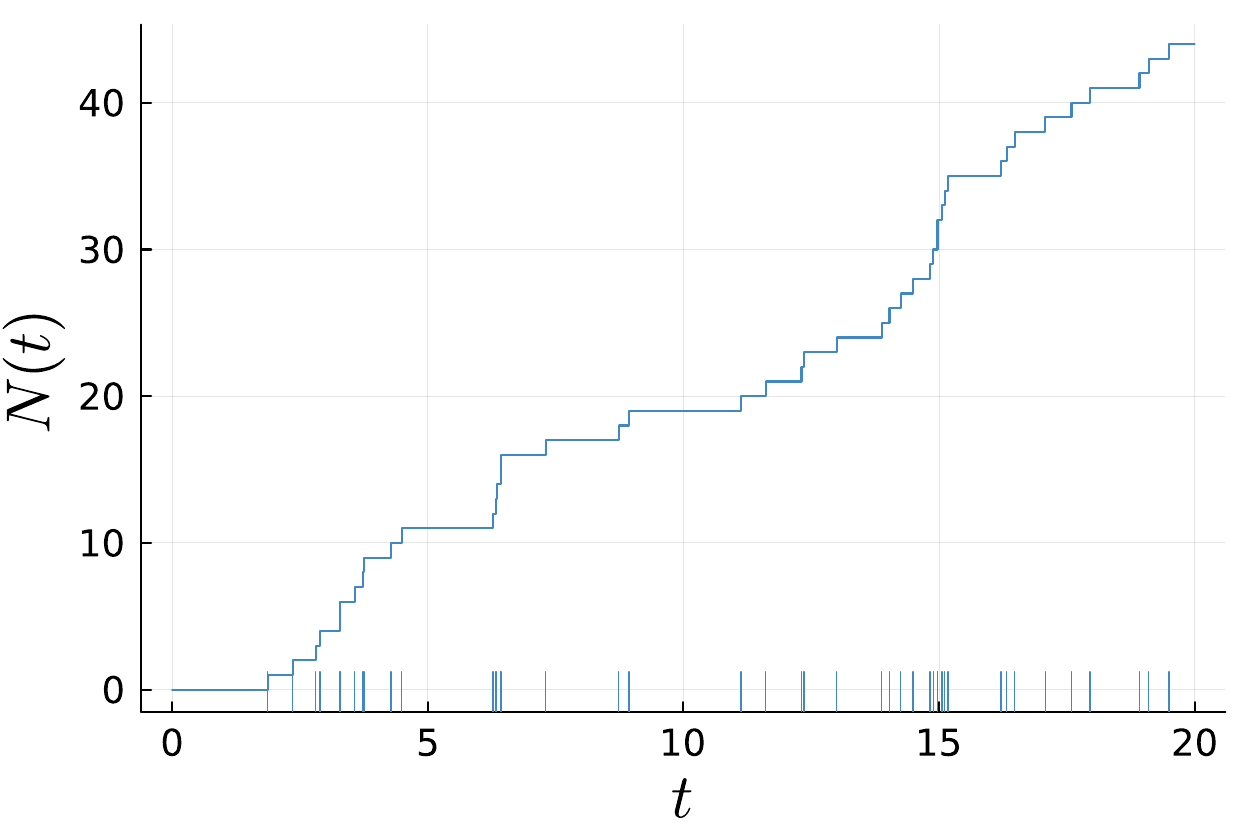}
    \label{fig:Nt}
\end{subfigure}
\vspace{-2.5em}
    \caption{Plot of intensity process $\lm(t)$ and counting process $N(t)$ for a simulated Hawkes process on $t\in[0, 10]$ with background rate $\nu(t)\equiv \nu = 1.0$, branching ratio $\eta = 0.6$ and offspring kernel $g(t) = e^{-t/0.5}/0.5$. 
    }
    \label{fig:examp}
\end{figure}

Let the strictly increasing sequence $\{\tau_i\}_{i\in \bb Z_+}\subset \bb R_+$ represent a realisation of a point process on the positive real line. Each element $\tau_i$ is interpreted as the occurrence time of the $i$\thh event after the initial time $t =  0$. Denoting by $\cc B(\cc X)$ the Borel $\sigma$-algebra over space $\cc X$, the associated counting process $N:\cc B(\bb R_+)\to\bb Z_+$ gives the number of events occurring on a measurable subset of the positive half line, formally,
\begin{align*}
    N(A) \ &= \ \sum_{i=1}^{\infty}\ind_A(\tau_i),\quad A\in \cc B(\bb R_+),
\end{align*}
where $\ind_A(\cdot)$ is the indicator function for set $A$. In particular, we use the notation $N(t)\, :=\, N(0,\, t]$ to represent the cumulative number of events from the origin to time $t$. Letting $\sigma(\cc X)$ denote the $\sigma$-algebra generated by the collection of sets $\cc X$, the history of the process is contained in the natural filtration $\{\cc F_t\}_{t\geq 0}$, where $\cc F_t =  \sigma\big(N(s): s\,\leq \, t\big)$. Letting $\cc F_{t-} =  \sigma\big(N(s): s\, <\, t\big)$, the Hawkes process model can be specified by the conditional intensity $\lm:\bb R_+\to\bb R_+$, defined as
\begin{align*}
    \lm(t) \ &:= \ \frac{\bb E\big[\dd N(t)\mid \cc F_{t-}\big]}{\dd t} \ = \ \nu(t) \ + \ \eta\int_{0}^{t-}g(t\, -\, s)\dd N(s).
\end{align*}
The \textit{background rate} $\nu(\cdot)$ is a strictly positive function that determines the baseline arrival rate of events, and the \textit{excitation kernel} $g(\cdot)$ is a probability density function on $\bb R_+$ that controls the shape and duration of the self-excitation effects. It is also known as the offspring density function since it specifies the birth time distribution of first-generation offspring due to an event. The \textit{branching ratio} $\eta$ is confined to the interval $[0,\, 1)$ to guarantee stability of the process and determines the expected number of first-generation offspring events triggered by any given event arrival. We assume that functions $\nu(\cdot)$ and $g(\cdot)$ are fully characterised by parameter vectors $\tht_\nu$ and $\tht_g$, respectively. The complete parameter of the Hawkes process is the $d$-dimensional vector $\theta =  (\tht_\nu,\, \eta,\, \tht_g)$, which is an element of the parameter space $\Tht\subset \bb R^d$. Figure~\ref{fig:examp} shows a short segment of the intensity process and counting process, respectively, of a simulated Hawkes process sample path.

\subsection{Interval Censoring and Likelihood}
When the Hawkes process is continuously observed up to time $t$, parameter estimates can be obtained by maximising the log-likelihood function, which is expressed (up to an additive constant independent of $\theta$) relative to the measure of a unit Poisson process as \citep[Proposition 7.2.III;][]{daleyIntroductionTheoryPoint2003a}
\begin{align*}
    \log L^{(c)}_t(\tht) \ &= \ \sum_{i=1}^{N(t)}\log \lm (\tau_i) \ - \ \int_0^{t}\lm(s)\dd s.
\end{align*}
Assume now that observations of $N$ are taken at $K\in \bb Z_+$ discrete time points $0 =  t_0 < t_1 < \dotsc < t_K$, with the censoring time $t_K$ equivalently denoted by $T$. The resulting observed data is the count sequence $n_{1:K} := (n_1,\dotsc, n_K)$, where each $n_k$ is the realised value of $N(t_{k-1},\, t_k]$. In the setting of discrete observations, the likelihood function is the analytically intractable joint probability
\begin{align*}
    L_K(\tht) \ &= \ \bb P_\tht\big(N(t_{k-1},\, t_k]\, =\, n_k,\, k\, =\, 1,\,\dotsc,\, K\big) \ =: \ p_\tht(n_{1:K}).
\end{align*}

\subsection{Neural Networks for Statistical Inference}\label{ssec:nn_inference}
Given the computational cost of accurately approximating the intractable likelihood, we pursue a likelihood-free approach, where a NN is trained to perform inference on the parameter ${\bs \tht}$ \citep{chonLinearNonlinearARMA1997, creelNeuralNetsIndirect2017}. This constitutes a NN regression problem, which we will now detail. For a complete introduction to feed-forward NNs, see Chapter 6 of \cite{Goodfellow-et-al-2016}, and for a useful review of NN techniques for parametric inference, see \cite{zammit-mangionNeuralMethodsAmortized2024}.

Recalling that our data is a sequence of counts $n_{1:K}\in \bb Z_+^K$, we first define the function $\bs s: \bb Z_+^K \to \bb R^s$, which computes an $s$-dimensional summary statistic from a given observation. Details of our choice of $\bs s$ for the present problem are given in Section~\ref{sec:sumstat}. Let ${\bs \tht}_q$ denote the marginal $q$-quantiles of the posterior distribution, $p({\bs \tht}\mid \bs s)$. The goal is to produce a single model that takes $\bs s$ as input, with the output targeting the vector $({\bs \tht}_\zeta,\, {\bs \tht}_{0.5},\,  {\bs \tht}_{1-\zeta})$, for $\zeta \in (0, 0.5)$. Naturally, our point estimate targets the posterior median, ${\bs \tht}_{0.5}$, with the upper and lower $\zeta$-quantiles defining the desired credible interval. The chosen model for this inference task is a fully connected, feed-forward NN, which consists of layers of nodes: an input layer, multiple hidden layers, and an output layer. The input layer is the summary statistic $\bs s\in \bb R^s$. Since we require three quantiles per element of ${\bs \tht}$, the output will be the estimate
$({\bs \hth}_\zeta,\, {\bs \hth}_{0.5},\,  {\bs \hth}_{1-\zeta})\in \bb R^{3d}$.

Suppose that a NN is specified with $L$ hidden layers, with $J_l$ nodes in layer $l\in \{1,\dotsc,L\}$. Each node in layer $l$, say $X^{(l)}_j$ for $j\in\{1,\,\dotsc,\, J_l\}$, is a multivariate, real-valued function. A given node receives input from \textit{all} nodes in the previous layer. The node passes a linear combination of these inputs through a non-linear \textit{activation function}, then transmits this information to the nodes in the next layer. Formally, we have 
\begin{align*}
    X^{(l)}_j\big(w^{(l)}_{1:J_{l-1},j},\, b_j^{(l)}\big) \ &= \ \phi_l\Big(\sum_{i=1}^{J_{l-1}}w^{(l)}_{i,j}X^{(l-1)}_i \ + \ b_j^{(l)}\Big),\quad l = 1,\dotsc,L,
\end{align*}
where $\phi_l$ is the activation function, each $w^{(l)}_{i,j}$ is a \textit{weight}, and $b_j^{(l)}$ is an additional constant called the \textit{bias} of the $l$\thh hidden layer. For our model, we use the Rectified Linear Unit \citep[ReLU;][]{nairRectifiedLinearUnits2010} activation for hidden layers. The output layer uses a linear activation function, which in our model produces the vector $\bs v\in \bb R^{3d}$. 

To finalise the NN design for our particular problem, output $\bs v$ is manually transformed to enforce monotonicity of the quantile estimates across each dimension. Drawing on the approach in \cite{sainsbury-daleLikelihoodFreeParameterEstimation2024}, we define the final outputs to be
\begin{align*}
    {\bs \hth}_{\zeta} \ &= \ \bs v_{1:d},\quad
    {\bs \hth}_{0.5} \ = \ \bs v_{1:d} \ + \ f(\bs v_{d+1:2d}),\quad
    {\bs \hth}_{1-\zeta} \ = \ \bs v_{1:d} \ + \ f(\bs v_{d+1:2d}) \ + \ f(\bs v_{2d+1:3d}),
\end{align*}
where here $f(\cdot)$ denotes an element-wise application of the softplus function. The NN thus always produces sensible outputs, and can be succinctly formulated as a function $F_{\bs w}:\bb R^s\to\, \bb R^{3d}$, with the vector $\bs w$ containing all weights and biases. 

The goal of training is to select a weight vector $\bs w^*$ that minimises the prediction error of the NN according to a specified loss function. For this task, a large sample of training data is required. First, a sample of training parameters ${\bs \tht}^{(1:M)}$ is drawn independently from a prior $\pi(\dd {\bs \tht})$ over $\Tht$. A discussion of effective priors is given in Section~\ref{ssec:prior}. Then, for each training parameter ${\bs \tht}^{(m)}$, $m = 1,\dotsc, M$, a sample path of the Hawkes process is simulated and aggregated to form $n^{(m)}_{1:K}$, from which the summary statistic $\bs s^{(m)} = \bs s\big(n^{(m)}_{1:K}\big)$ is computed. By producing the training sample as described, the NN is trained to approximate the statistical relationship between the parameters of the Hawkes process and the observed data without needing reference to the likelihood function of the Hawkes process. Since the Hawkes process can be simulated in linear time, generating the training sample is highly efficient.

As is detailed in \cite{sainsbury-daleLikelihoodFreeParameterEstimation2024}, the NN that results from training relative to loss function $\ell(\bs w)$ will target the classical Bayes estimator associated with $\ell$. For instance, using mean-squared error loss will target the posterior mean of $p({\bs \tht}\mid \bs s)$. To target the quantiles and median of $p({\bs \tht}\mid \bs s)$, we follow \cite{fisherDeepLearningMarginal2024} and \cite{sainsbury-daleNeuralBayesEstimators2025} by making use of the quantile loss function, defined as
\begin{align*}
    \ell_q(\bs w;{\bs \hth}) \ &= \  \sum_{m=1}^M \sum_{i=1}^d L_\zeta\big({\bs \tht}^{(m)}_i,\, {\bs \hth}_i \big),
\end{align*}
where, for $q\in (0, 1)$,
\begin{align*}
    L_q({\bs \tht}_i, {\bs \hth}_i) \ &= \ ({\bs \hth}_i\, -\, {\bs \tht}_i)(\ind\{{\bs \hth}_i > {\bs \tht}_i\}\, -\, q).
\end{align*}
Intuitively, $L_q$ penalises NN predictions such that the optimal estimator misses above the true parameter at a rate of $q$. The complete loss function applied to the outputs of the NN is therefore taken to be
\begin{align*}
    \ell \big(\bs w; ({\bs \hth}_\zeta,\, {\bs \hth}_{0.5},\,  {\bs \hth}_{1-\zeta})\big) \ &= \ \ell_\zeta(\bs w; {\bs \hth}_{\zeta}) \ + \ \ell_{0.5}(\bs w; {\bs \hth}_{0.5}) \ + \ \ell_{1-\zeta}(\bs w; {\bs \hth}_{1-\zeta}),
\end{align*}
which equally weights the tasks of estimating the median and upper and lower $\zeta$-quantiles. Our models are trained using the ADAM algorithm \citep{kingmaAdamMethodStochastic2014}, and we implement early stopping to prevent overfitting. By designing the NN as described and utilising the stated loss function, the trained model jointly produces a point estimate of ${\bs \tht}$ and marginal credible intervals for each dimension of ${\bs \tht}$, enabling fully amortised parametric inference.

\section{Summary Statistic and Prior Distribution}\label{sec:sumstat}
In this section, we detail our choice of summary statistics for the discretely observed Hawkes process. We then show how these can be extended to settings with unequally sized aggregation windows and/or time-varying background rates. We also give some practical guidelines for designing a prior distribution over the parameter space $\Tht$.

\subsection{Basic Summary Statistic}\label{ssec:sumstat}
The quality of the NN estimates relies on selecting a summary statistic that is sensitive to small changes in the parameter. Experimentation found that standard summary statistics used in \cite{creelNeuralNetsIndirect2017} such as mean, variance and auxiliary regressions, are not effective at identifying the parameters of the excitation kernel when applied to the interval censored Hawkes process. Additionally, we desire a summary statistic that is computable in linear time, to facilitate the rapid generation of training samples. A final criterion for the ideal summary statistic is that it is of the smallest dimension that allows for identification of the parameters, as this reduces the size of the corresponding NN, improving training speed and performance.

In this section, we propose a novel summary statistic that is constructed from two misspecified models. It satisfies the properties outlined above, with the quality of the resulting NN estimates demonstrated in Section~\ref{sec:nn_simstudy}. Importantly, the principle upon which the summary statistic is formulated can feasibly be generalised to other processes under incomplete information. For now, we assume that the background rate is constant and the censoring intervals are equal in width. These assumptions will be relaxed in Section~\ref{ssec:timevary}.

\subsubsection{Single Imputation Estimate}
Single imputation is an estimation technique designed for settings with missing data (see Chapter 4 of \cite{littleStatisticalAnalysisMissing2020} for a detailed discussion), which we apply to the interval censored Hawkes process to obtain a useful summary statistic. Let $N_k = N(t_k)$ be the total number of events occurring up to the $k$\thh observation time. An imputed sample path $\tau^\mathrm{imp}_{1:N_K}$ is constructed by setting
\begin{align*}
    \tau^\mathrm{imp}_{N_{k-1}+i} \ &= \ t_{k-1} \, +\, \frac{i}{n_k+1},\ i\, =\, 1,\dotsc,n_k.
\end{align*}
The imputation estimate, $\hth^\mathrm{imp}$, is the MLE obtained from the imputed sample path.

When the Hawkes process is specified with an exponential offspring distribution, the intensity of the process is Markovian, allowing the MLE to be computed in linear time. In this case, $\bs s(n_{1:K}) = \hth^\mathrm{imp}$ is a minimum-dimension summary statistic that is rapid to generate and is highly sensitive to changes in all parameters. As we will illustrate in Section~\ref{sec:nn_simstudy}, the resulting NN estimates demonstrate a similarly good performance as the PMMH estimator. Importantly, the placement of latent event times makes no attempt to accurately capture the true structure of events from the Hawkes process, thus avoiding the detailed constructions used in \cite{shlomovichUnivariate2022} and \cite{schneiderEstimationSelfexcitingPoint2023}. 

Though the imputation estimate for non-exponential kernels similarly provides a highly effective, minimum-dimension summary statistic, it requires quadratic computational time to compute the MLE. This is impractical without access to a high-performance computing cluster, given the large training samples that are needed for training a NN. For this reason, we purposefully fit a misspecified Markovian Hawkes process to the imputed data. The imputation estimate of the exponential excitation kernel can be interpreted as an estimator of the mean offspring waiting time, which remains highly sensitive to the parameters of the offspring distribution. To complete the summary statistic, we fit an autoregression on the observed count data, described in the next section.

\subsubsection{Negative Binomial Autoregression}
To supplement $\hth^\mathrm{imp}$ in the case of a non-exponential excitation kernel, we also fit a Negative Binomial autoregression (NBAR) to the observed count data. A NBAR($p$) model, with $p\in \bb Z_+$ denoting the number of lagged covariates, assumes that $n_k\mid n_{k-p:k-1}, \phi_k  \sim \mathrm{Poi}(\mu_k\phi_k)$
where $\mu_k > 0$ is a rate parameter and $\phi_k\,\siid\, \mathrm{Gamma}(\delta,\, \delta)$ is an unobserved random variable. This is a generalisation of the Poisson AR model. Integrating out $\phi_k$ yields the conditional distribution $n_k \mid n_{k-p:k-1}  \sim  \mathrm{NegBinom}\big(\delta/(\delta\, +\, \mu_k), \delta\big)$. 
We use the typical logarithmic link function to model the rate, which assumes that $\log \mu_k = \gamma_0\, +\, \sum_{i=1}^p\gamma_{i}n_{k-i}$.
The estimates $\hat\gamma_{0:p}$ of $\gamma_{0:p}$ are obtained via MLE, and comprise the next $p+1$ dimensions of the summary statistic. The NBAR estimates capture the effect of recent event counts on the observation in a given window; hence, they are sensitive to the distribution of waiting times to offspring events. The number of lags, $p$, is flexible and should be chosen to suit the specific problem. Details of the impact of varying $p$ on the performance of the estimator, along with some practical recommendations for selecting $p$, are given in Section~\ref{ssec:nlags}.

We also obtain an estimate $\hat\delta$, which is the final element in the summary statistic. The parameter $\delta$ is referred to as the \textit{dispersion parameter}, and quantifies the level of overdispersion of the data relative to a Poisson process. In particular, the conditional variance of the count data is $\mathrm{Var}(n_k\mid n_{k-p:k-1}) =  \mu_k + \mu_k^2/\delta$. The final summary statistic in the case of a non-exponential excitation kernel is $\bs s(n_{1:K}) =  \big(\hth^\mathrm{imp},\, \hat \gamma_{0:p},\, \hat \delta\big)$.

\subsubsection{Motivations for the use of Summary Statistics}
As is common in many other works \citep{lenziNeuralNetworksParameter2023, sainsbury-daleLikelihoodFreeParameterEstimation2024, sainsbury-daleNeuralBayesEstimators2025}, it is possible to train the NN to accept the complete sequence $n_{1:K}$ as input. Given that the data is an integer time series, this is likely best achieved using a Recurrent Neural Network (RNN). The benefit of this approach is that the RNN attempts to select optimal summaries of the data during training, which may improve performance. Instead, we have followed the work of \cite{creelNeuralNetsIndirect2017}, designing a summary statistic that fits within the indirect inference framework, which provides some important advantages for this problem.

Our summary statistic $\bs s(n_{1:K})$ is the estimator of an auxiliary parameter whose dimension is higher than or equal to that of the target parameter $\theta$, although for the exponential case, the dimensions of $\bs s(n_{1:K}) = \hth^\mathrm{imp}$ naturally correspond with the counterpart dimensions of $\tht$. The classical Pitman-Koopman-Darmois theorem \citep{pitmanSufficientStatisticsIntrinsic1936, koopmanDistributionsAdmittingSufficient1936} states that, under technical conditions pertaining to continuity and support of the canonical statistic, a sufficient statistic of fixed dimension for a parametric model exists if and only if that model is in the exponential family. Given that the parametric model applied in our work, $p_\tht(n_{1:K})$, is analytically intractable, finite dimensional sufficient statistics are not expected to exist. However, when an auxiliary model is regular and estimated using MLE, the auxiliary parameters are known to suffer from very little information loss in a range of models \citep{drovandiBayesianIndirectInference2015}. Though proof of this claim is difficult when working with intractable models \citep{drovandiBayesianIndirectInference2015}, we endeavour to explore the sufficiency of $\bs s$ numerically in Section~\ref{sec:nn_simstudy} through comparison to PMMH, which is a statistically sound technique that utilises the complete data set.

The primary advantage of our chosen summary statistic is that its dimension, $s\in \bb Z_+$, is independent of $K$, the number of observations. As such, NN estimation can be conducted on sample paths of different lengths using the same trained model, without requiring padding techniques \citep{creelNeuralNetsIndirect2017}. This dramatically improves the amortisation of the method, for instance, in settings where one collects more data over time. The estimation procedure's ability to handle data with varying censoring times is justified by the asymptotic stability of the imputation scheme and the NBAR estimate. Since the summary statistic appears to be $\sqrt{K}$-consistent (Appendix, Section C), one can safely apply a trained NN to any sample path with a censoring time sufficiently large to ensure that the imputation estimates are close to convergence, which can be checked numerically. The quantile estimation procedure for paths of differing lengths requires a rescaling step, which is detailed in Section D of the appendix. 

A key difference between our summary statistic and the many used by \cite{creelNeuralNetsIndirect2017} is that a natural correspondence exists between $\hth^\mathrm{imp}$ and $\tht$. This allows us to use dense NNs with a fairly small number of nodes, since we only need to approximate a relatively low-dimensional mapping, which appears to be smooth in the exponential kernel case (Section A, Appendix). As a result, convergence during training is rapid, compared to training a RNN on the full event count sequence. Additionally, unlike the complex path reconstruction schemes used in \cite{shlomovichUnivariate2022} and \cite{schneiderEstimationSelfexcitingPoint2023}, the imputation estimate takes very little expert knowledge to design and is straightforward to implement. The use of imputation estimates with NNs may be a generally useful inference technique in settings with incomplete data and warrants further exploration.

\subsection{Non-Constant Interval Censoring and Time-Varying Background Rates}\label{ssec:timevary}
In certain cases, the event times of a point process are subject to aggregation over censoring intervals that differ in size. One such example is COVID-19 case numbers across Australia, whereby each state moved from daily infection count reporting to weekly reporting in September 2022, after a cost assessment and consultation with health officials \citep{australianbroadcastingcorporationCOVID19StatisticsMove2022}. Additionally, the Hawkes process can be specified with a time-varying background rate function, $\nu(t)$, which is relevant in a variety of applications, for instance, seasonally fluctuating infectious disease counts. In both cases, some minor adjustments to the summary statistic are required.

\subsubsection{Non-Constant Interval Censoring}
The observation times $\{t_k\}_{k= 0}^K$ may arise stochastically or deterministically. We require only that they are known to the observer and are independent of the process $N(t)$. In this setting, the imputation estimates are obtained identically to the case of constant interval censoring. However, for the NBAR($p$) estimates, we work in a similar setting where $N_k\mid N_{k-p:k-1},\phi_k  \sim  \mathrm{Poi}(\mu_k\phi_k)$
but the autoregression is now performed on the time-standardised rate of event arrivals according to
\begin{equation*}
     \log(\mu_k/\Delta_k) \
     = \ \gamma_0 \ + \ \sum_{i=1}^p\gamma_{i}(n_{k-i}/\Delta_{k-i}).
\end{equation*}
Modelling $\mu_k/\Delta_k$ accounts for the fact that $n_k$ is observed over an interval of length $\Delta_k$, while using terms $n_{k-i}/\Delta_{k-i}$ as regressors normalises each lagged term to the same scale.

\subsubsection{Time-Varying Background Rate}
Recall that the rate function $\nu(\cdot)$ is assumed to depend on a vector of parameters $\tht_\nu$. The imputation estimate can therefore be obtained as in the case of a constant baseline. Suppose for now that the rate function is known. Defining the term $V_k = \int_{t_{k-1}}^{t_k}\nu(s)\dd s$, 
the NBAR($p$) estimates can be obtained in the same way as with unequal censoring intervals by modelling the mean via
\begin{align*}
    \log\mu_k \ &= \ \log V_k + \ \gamma_0 \ + \ \sum_{i=1}^p\gamma_{i}(n_{k-i}/V_{k-i}).
\end{align*}
This accounts for the changing volume of background event arrivals over each period. We use the piecewise approximation $V_k \approx \nu\big(t_{k-1} + \Delta_k/2\big)\Delta_k$ to the volume term,
which works well in practice and introduces minimal error when the variation of $\nu(\cdot)$ over each interval is small. Since $\nu(\cdot)$ contains the unknown parameters $\tht_\nu$, in the offset term we replace $V_k$ with $\hat\nu^\mathrm{imp}\big(t_{k-1}\, +\, \Delta_k/2\big)\Delta_k$
instead, where $\hat\nu^\mathrm{imp}(\cdot)$ denotes the function $\nu(\cdot)$ with the  $\tht_\nu$ replaced by $\hth^\mathrm{imp}$. Though this is a rough approximation, the NN is still effective at discerning the underlying parameters from the summary statistic. 

\subsection{Prior Distribution}\label{ssec:prior}
The parameter space of the Hawkes process, $\Tht$, has $\eta\in [0,\, 1)$, with all other parameters being subject to constraints based on the choice of $\nu(\cdot)$ and the offspring kernel. Simulation experiments have shown that the results of the NN estimator are not greatly impacted by the choice of prior, so the priors used in this work are chosen for other practical considerations. A prior covering a large subset of $\Tht$ should be preferred to improve the generalisability of the trained model.

Since $\eta\in [0, 1)$ the sample is drawn from a standard normal on the $\logit$ scale, that is, $\logit\big(\eta^{(1:M)}\big)\siid \cc N(0,\, 1)$. This is chosen instead of the $U(0,1)$ distribution for minor advantages in computational speed. Since the total number of events to time $T$ satisfies $\frac{N(T)}{T}\to \frac{\nu}{1-\eta}$, sampling on the logit scale as described shifts the mass of the prior towards the centre of the interval $[0,1)$, reducing the number of extreme event counts in the training sample.

It is often the case that elements of the parameters $\tht_\nu$ and $\tht_h$ will be restricted to $\bb R_+$, for which Gamma or Log-normal priors are common in the Bayesian literature. Instead, we use an alternative prior that places significant mass near the lower bound of $0$. Firstly, note that the softplus function is defined by $f(x) =  \log\big(1\, +\, e^x\big)$. For a representative parameter $\alpha\, >\, 0$, we sample $f^{-1}\big(\alpha^{(1:M)}\big)\, \siid \, \cc N\big(\mu_\alpha,\, \sigma_\alpha^2\big).$ We henceforth refer to this distribution as the \textit{inverse softplus normal} (ISN) distribution. 
The ISN prior can be made to place more mass near $0$ than a Gamma or Log-normal distribution with equivalent mean and variance, which can improve the performance of the NN quantile estimator for small parameter values. When working with certain time-varying background rates (such as the spline in Section~\ref{sec:app_infec_dis}), elements of $\tht_\nu$ may be allowed to take negative values. In this case, a normal distribution is selected as the prior.

\section{Simulation Study}\label{sec:nn_simstudy}
In this section, we assess the quality of the NN estimator on various simulated sample paths. The NN estimator is compared to competitor methodologies in the literature, and we also illustrate the performance of the method with different lag sizes for the NBAR estimates, $p$, as well as time-varying background rates.


\subsection{Initial Experiments}

\subsubsection{Exponential Kernel}\label{ssec:nn_simstudy_pmmh}
For an exponential excitation kernel, the uniform imputation estimate can be used as a high-quality summary statistic that is fast to obtain. We use the PMMH estimator developed in \cite{chenEstimatingHawkesProcess2025} as a benchmark of the extant methods, implemented with the ordered uniform proposal suggested in \cite{lambeFittingMultivariateHawkes2025}, due to its numerical performance improvements. The data are simulated to a censoring time $T = 1,\!000$, with varying levels of aggregation, $\Delta\, >\, 0$. A training sample of size $M = 100,\!000$ is drawn from the prior described in Section~\ref{ssec:prior}, with $\nu^{(1:M)}$ and $\beta^{(1:M)}$ drawn independently from the $\mathrm{ISN}(5, 9)$ distribution. For this experiment, $J = 500$ test sample paths are generated from the true parameter and then estimated, with the results presented in Table~\ref{tab:pmmh}. The reported estimates (Est) are the respective mean estimates for each estimation procedure, along with their respective standard errors (SE). The posterior $95\%$ marginal credible intervals are also approximated for each sample path using the quantile NN method described in Section~\ref{ssec:nn_inference}, with the proportion of intervals containing the associated parameter reported in row $\mathrm{CP}$. The associated values for the PMMH method are computed as in \cite{chenEstimatingHawkesProcess2025}.

\begin{table}
    \centering
    \caption{NN and PMMH parameter estimates with exponential kernel, $T = 1,\!000$. NN model: two hidden layers with 64 and 32 nodes, respectively.}
    \setlength{\tabcolsep}{2.75pt}
    \begin{tabular}{lccccclccccc}\toprule
     & & & $\nu$ &$\eta$ &$\beta$ & & & & $\nu$ &$\eta$ &$\beta$ \\
     \midrule
     & & & $2.0$ & $0.6$ & $2.0$ & & & & $2.0$ & $0.6$ & $2.0$ \\ 
    \midrule
\multirow{6}{*}{$\Delta\, =\, 0.1$} & \multirow{3}{*}{NN} & Est & 2.046 & 0.591 & 2.067  &\multirow{6}{*}{$\Delta\, =\, 0.5$} & \multirow{3}{*}{NN} & Est & 2.062 & 0.589 & 2.037 \\
&& SE  & 0.214 & 0.046 & 0.308  &&& SE & 0.224 & 0.044 & 0.309 \\
&& CP & 0.958 & 0.958 & 0.934 &&& CP & 0.956 & 0.958 & 0.932 \\
\cmidrule(lr){2-6} \cmidrule(lr){8-12}
& \multirow{3}{*}{PMMH}& Est & 2.002 & 0.600 & 2.062
 && \multirow{3}{*}{PMMH} & Est & 2.005 & 0.600 & 2.056\\
&& SE & 0.215 & 0.045 & 0.314  &&& SE & 0.218 & 0.046 & 0.314\\
&& CP & 0.960 & 0.952 & 0.939 &&& CP & 0.958 & 0.960 & 0.942\\
\midrule

\multirow{6}{*}{$\Delta\, =\, 1.0$} & \multirow{3}{*}{NN} & Est & 2.034 & 0.592 & 2.094
 & \multirow{6}{*}{$\Delta\, =\, 5.0$} & \multirow{3}{*}{NN} & Est & 2.062 & 0.586 & 2.100\\
 && SE  & 0.214 & 0.044 & 0.358 &&& SE & 0.259 & 0.051 & 0.510 \\
&& CP & 0.956 & 0.958 & 0.952  &&& CP & 0.952 & 0.938 & 0.976 \\
\cmidrule(lr){2-6} \cmidrule(lr){8-12}
& \multirow{3}{*}{PMMH} & Est & 2.011 & 0.598 & 2.050 && \multirow{3}{*}{PMMH} & Est & 2.162 & 0.567 & 1.870\\
&& SE & 0.241 & 0.050 & 0.330 &&& SE & 0.468 & 0.096 & 0.862 \\
&& CP & 0.960 & 0.960 & 0.948 &&& CP & 0.904 & 0.908 & 0.920 \\
    \bottomrule
\end{tabular}
    
    \label{tab:pmmh}
\end{table}

Both methods exhibit very little empirical bias, particularly for small $\Delta$ values. The magnitude of the standard errors is generally comparable, though notably smaller for the NN estimator in the case of $\Delta = 5.0$. It may be possible to remove this discrepancy by tuning the PMMH algorithm by adjusting the number of particles and the step size of the proposal distribution, though this is a practical challenge in implementing the PMMH algorithm \citep{chenEstimatingHawkesProcess2025}. The coverage probabilities are also well calibrated for both methods. The PMMH chain has the full information posterior $p(\tht \mid n_{1:K})$ as its stationary distribution \citep{chenEstimatingHawkesProcess2025}. The closeness of the NN estimator to the PMMH estimator for both point estimation and quantile estimation is therefore strong numerical evidence that the proposed summary statistic is highly informative. 

\begin{figure}[h]

\begin{subfigure}{0.32\textwidth}
\centering
    \includegraphics[width = \textwidth]{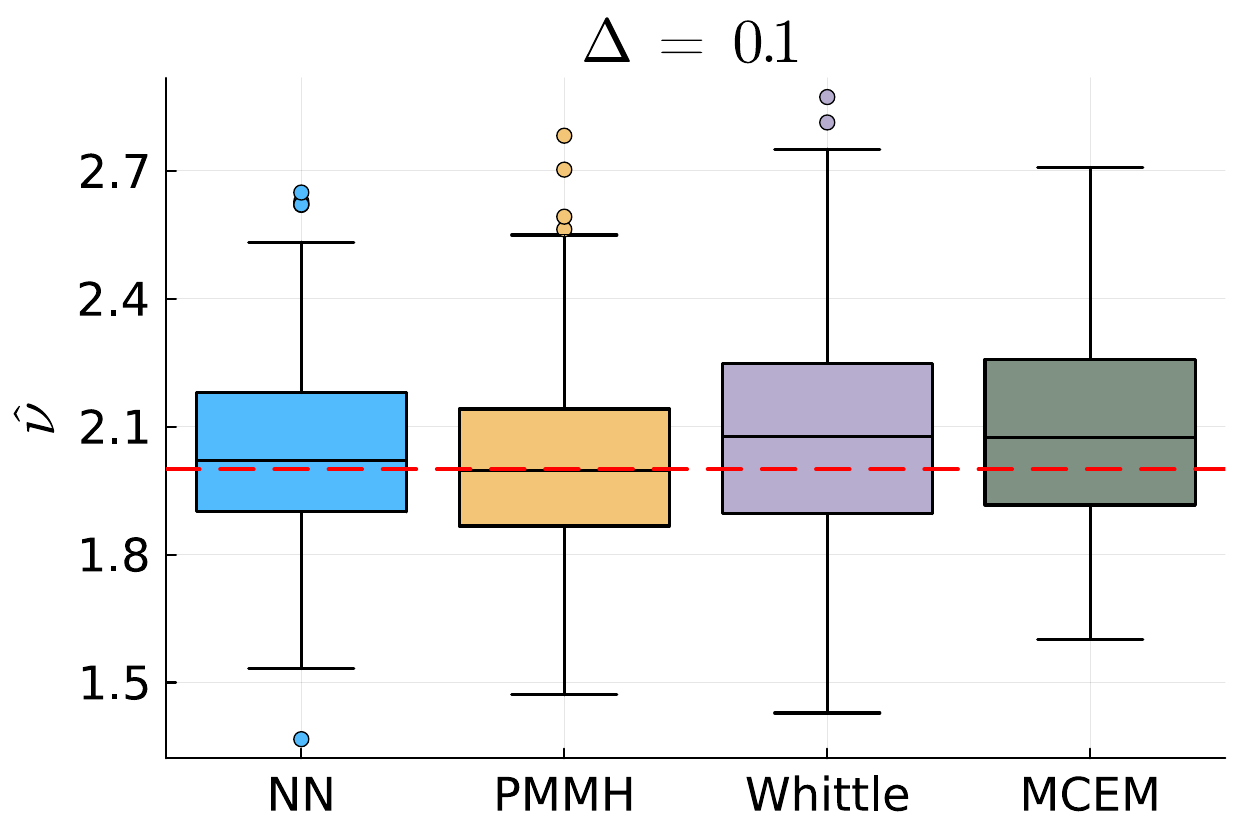}
    \label{fig:nu_0.1}
\end{subfigure}
\hfill
\begin{subfigure}{0.32\textwidth}
    \centering
    \includegraphics[width = \textwidth]{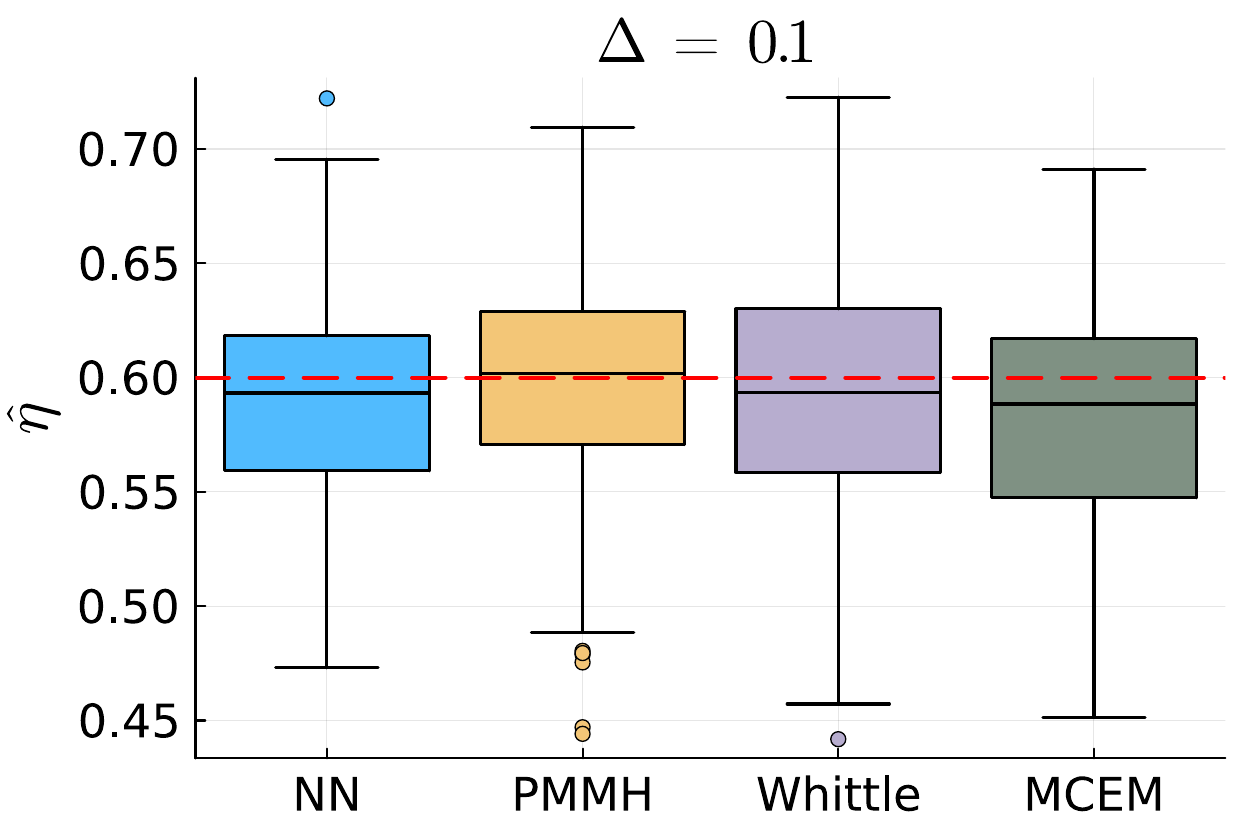}
    \label{fig:eta_0.1}
\end{subfigure}
\hfill
\begin{subfigure}{0.32\textwidth}
    \centering
    \includegraphics[width = \textwidth]{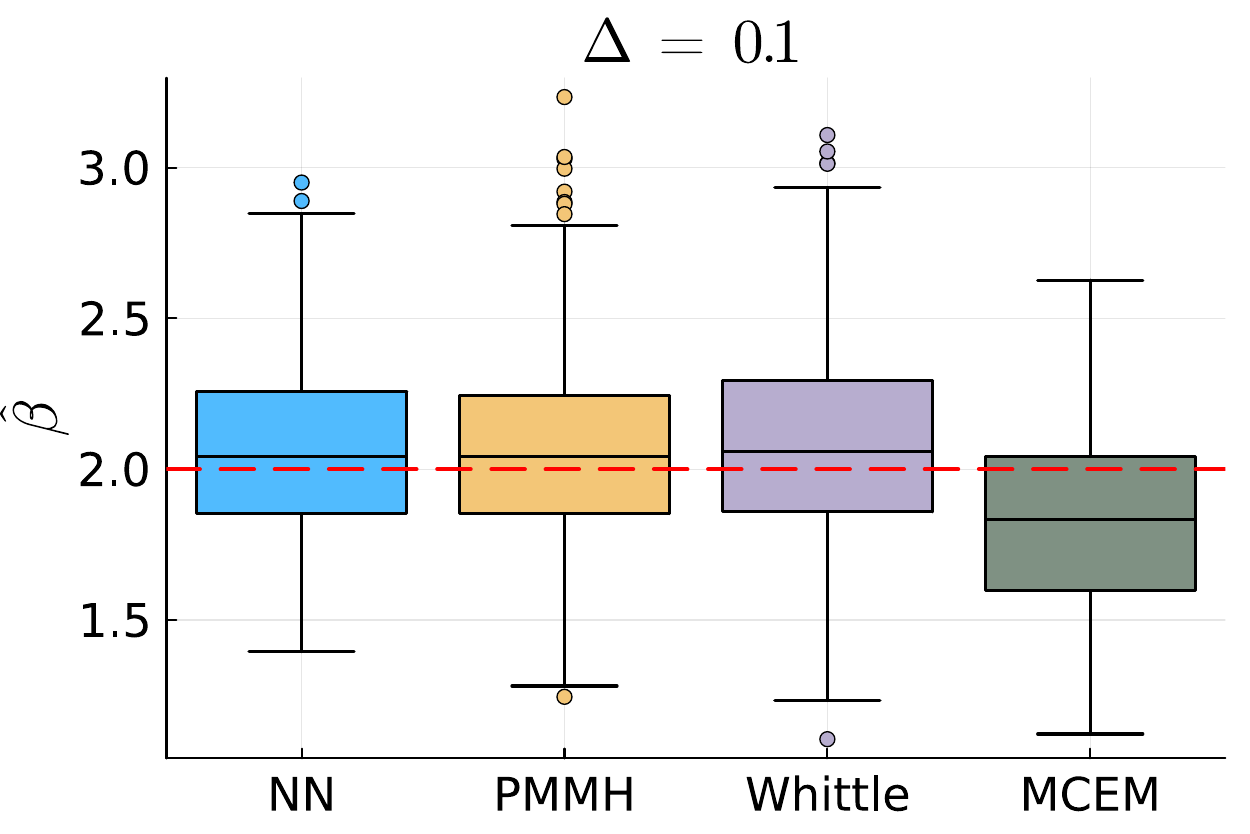}
    \label{fig:beta_0.1}
\end{subfigure}
\par\vspace{-1.5em}
\begin{subfigure}{0.32\textwidth}
\centering
    \includegraphics[width = \textwidth]{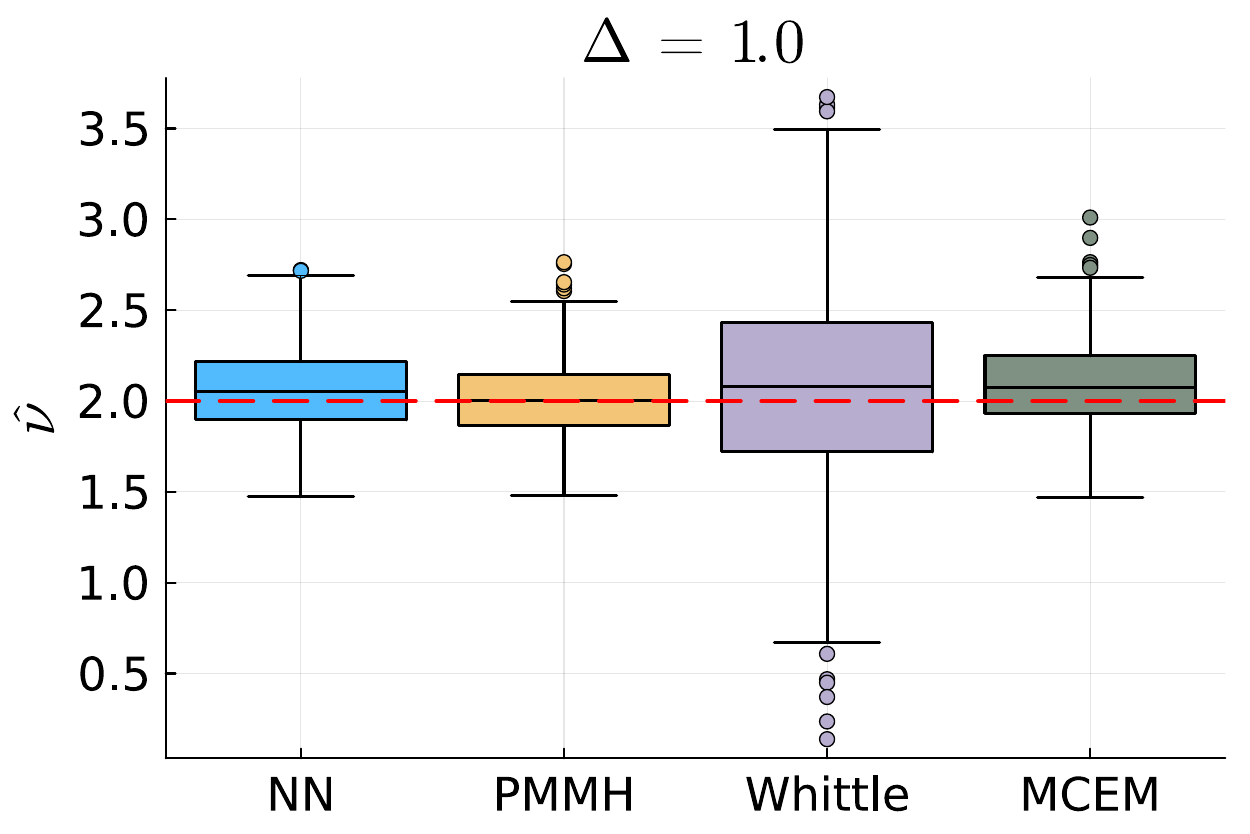}
    \label{fig:nu_1.0}
\end{subfigure}
\hfill
\begin{subfigure}{0.32\textwidth}
    \centering
    \includegraphics[width = \textwidth]{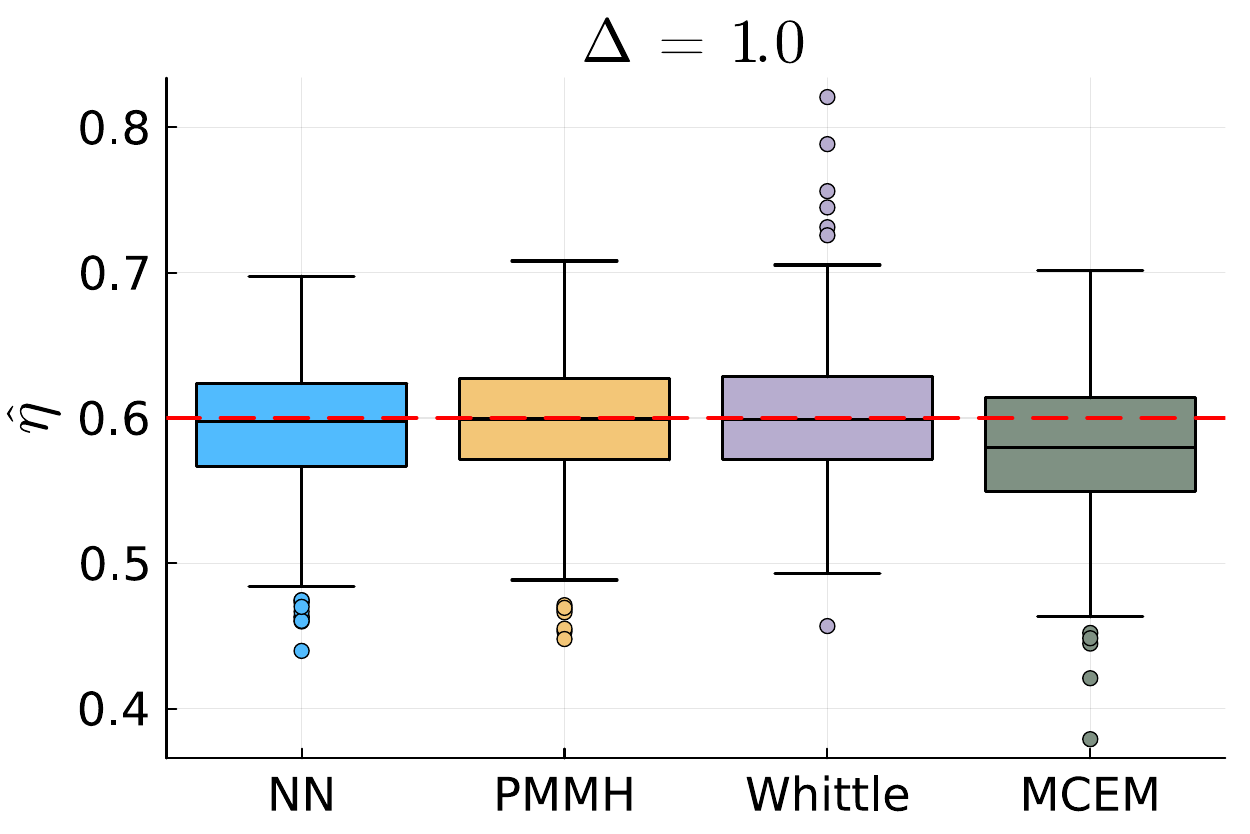}
    \label{fig:eta_1.0}
\end{subfigure}
\hfill
\begin{subfigure}{0.32\textwidth}
    \centering
    \includegraphics[width = \textwidth]{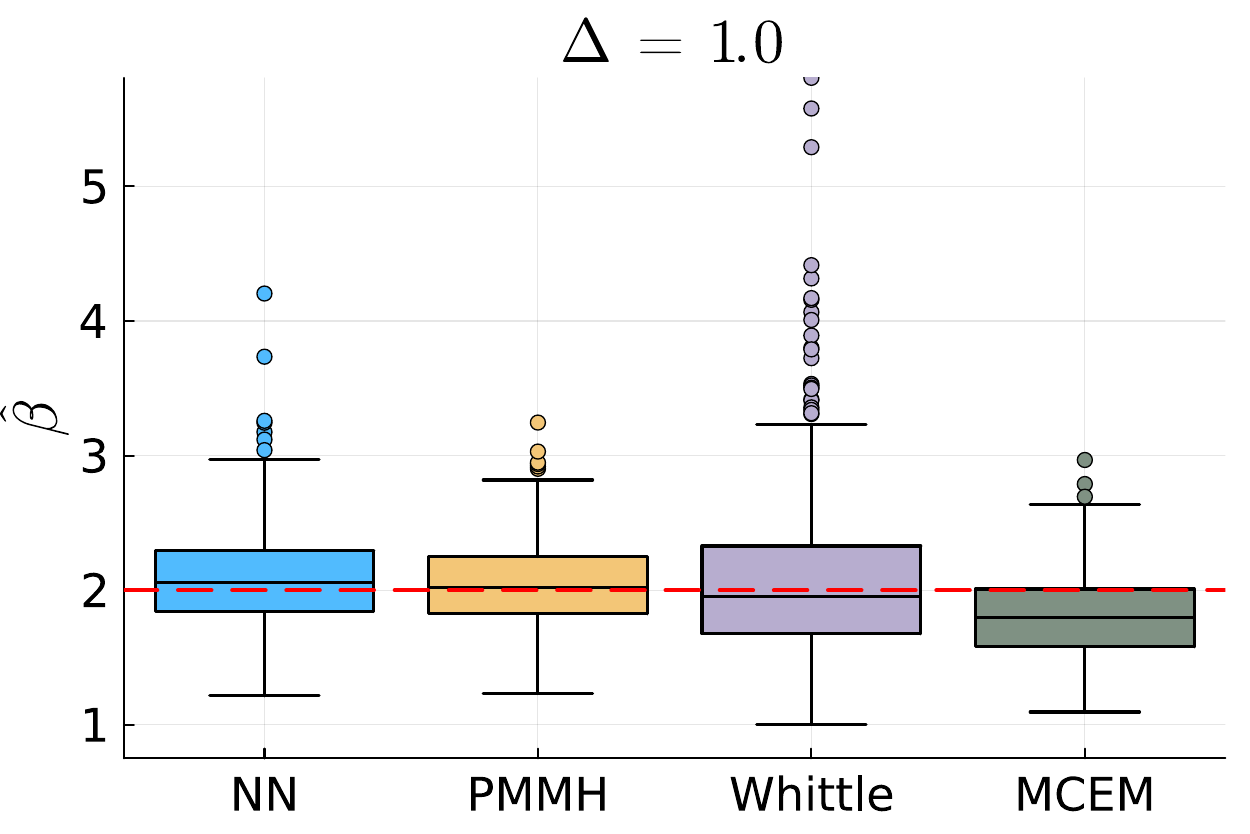}
    \label{fig:beta_1.0}
\end{subfigure}
\par\vspace{-2.5em}
\caption{Boxplots of estimates from NN, PMMH, Whittle and MCEM methods. Test paths are simulated to censoring time $T=1,\!000$ from parameter $\tht = (2.0, 0.6, 2.0)$ with an exponential kernel. Aggregation levels $\Delta = 0.1$ and $\Delta = 1.0$ are presented.}
\label{fig:boxplots}
\end{figure}

Figure~\ref{fig:boxplots} presents a visual comparison of the results in Table~\ref{tab:pmmh} to the Whittle estimator \citep{cheyssonSpectralEstimationHawkes2022} and MCEM estimator \citep{shlomovichUnivariate2022} in the cases of $\Delta = 0.1$ and $\Delta = 1.0$. The Whittle estimator has greater variability than the NN and PMMH estimators, and the MCEM estimator has significant bias. In Section A of the Appendix, we apply the NN estimator across a range of parameter values for the $\Delta = 0.1$ case to better understand the behaviour of the estimator. The trained model is a smooth, approximately linear function of the imputation estimate.

\subsubsection{Non-Exponential Kernel}\label{ssec:gamma_kern}
We now apply the NN estimation procedure to a Hawkes process with a $\mathrm{Gamma}(1.5,\, 0.25)$ excitation kernel. The results are displayed in Table~\ref{tab:gam}, for differing levels of aggregation. As discussed in Section~\ref{sec:sumstat}, we now include the $\mathrm{NBAR}(p)$ estimates in the summary statistic to enable identification of the parameters. It is challenging to present a fair comparison over different $\Delta$ values; for a given $\Delta$, using $p$ lags only allows the summary statistic to capture the impact of counts to within $p\Delta$ of each observation. For this particular experiment, we select $p_\Delta =  1/\Delta$ for each $\Delta$ value, so that the same duration of sample path history is captured in each experiment. The impact of the choice of $p$ will be explored in the next section. The NN estimates exhibit minimal bias, with the quantile estimates being well calibrated to the targeted coverage of $95\%$. A natural increase in standard error concurrent with an increase in $\Delta$ is also observed along the $\alpha$ and $\beta$ dimensions, as a larger aggregation window reduces the information carried about the offspring kernel in the count data. The same behaviour is not observed for the estimates of $\nu$ and $\eta$, as the presented $\Delta$ still yield highly informative data. It is clear that the NBAR($p$) estimates are capable of identifying the parameters of a non-exponential offspring density.

\begin{table}[ht]
    \centering
        \caption{NN estimates with Gamma kernel, $T =  1,\!000$. NN model: two hidden layers with 128 and 64 nodes, respectively.}
    \begin{tabular}{lccccccccccc}\toprule
     & & $\nu$ &$\eta$ & $\alpha$ &$\beta$ & && $\nu$ &$\eta$ & $\alpha$ &$\beta$\\
     \midrule
     & & $2.0$ &$0.6$ & $1.5$ &$0.25$ & && $2.0$ &$0.6$ & $1.5$ &$0.25$\\ 
\midrule
\multirow{4}{*}{\shortstack{MLE \\ ($\Delta=0$)}}
 & Est & 2.005 & 0.599 & 1.516 & 0.250 & \multirow{3}{*}{$\Delta=0.1$}& Est &  1.972 & 0.600 & 1.527 & 0.252 \\ 
 & SE &  0.104 & 0.022 & 0.131 & 0.031 & & SE  & 0.123 & 0.023 & 0.169 & 0.038 \\
& CP & 0.953 & 0.954 & 0.950 & 0.936 & & CP & 0.976 & 0.956 & 0.954 & 0.960 \\
   \midrule

\multirow{3}{*}{$\Delta\, =\, 0.2$} 
& Est & 2.011 & 0.598 & 1.575 & 0.236 & \multirow{3}{*}{$\Delta\, =\, 0.5$} 
& Est & 2.009 & 0.591 & 1.660 & 0.238 \\ 
& SE  & 0.107 & 0.022 & 0.172 & 0.031 && SE  & 0.098 & 0.023 & 0.340 & 0.060 \\
& CP & 0.958 & 0.946 & 0.934 & 0.954 && CP & 0.966 & 0.916 & 0.958 & 0.978 \\

\bottomrule
\end{tabular}

    \label{tab:gam}
\end{table}
\subsubsection{Comparison of NN and PMMH Estimation}
One major advantage of the NN estimation procedure over the benchmark PMMH estimator is that it can accurately estimate the parameters of non-exponential excitation kernels from interval censored data with minimal increase in computational time. On the other hand, PMMH estimation is much slower when applied to non-exponential kernels, as the Markov property of the intensity cannot be leveraged. 

Consider the case of $\Delta=0.1$ with $\mathrm{Gamma(1.5, 0.25)}$ kernel. Producing a NN estimation incurs two main sources of computational cost: producing the training sample, and training the NN. Training samples were produced using \texttt{Julia} in batches of $10,\!000$ on an Intel Xeon Platinum 8532Y system with 16 CPU cores. Each batch required approximately 2 minutes to run. This cost is of course amortised, and batches can be produced in parallel. One is therefore limited only by access to computational resources. The NN is trained in \texttt{Python} on a 12\thh Gen Intel Core i7-1255U processor, requiring approximately 5 to 10 minutes in total. The training speed may increase depending on the choice of batch size or complexity of the network, though our results do not vary specifically with changes in these factors. Once the NN is trained, estimates are obtained on the order of microseconds.

In comparison, on the same 16 CPU cores as with the training sample, the PMMH algorithm for this problem requires 6 to 8 hours per $1,\!000$ iterations. Therefore, a single estimate using a modest selection of $10,\!000$ iterations will require approximately 60 to 80 hours. The SMC algorithm for likelihood estimation can be run in parallel across particles \textit{within} a censoring interval. However, the computations for each particle must be completed before moving to the subsequent censoring interval. The PMMH procedure therefore cannot be accelerated arbitrarily with increasing computational resources, as can be done with the production of training samples for the NN estimation.

For the exponential Hawkes process, production of the training samples for NN estimation incurred the same computational cost. The reduced dimension of the summary statistic means that the training time reduces to approximately 2 minutes. Each individual PMMH procedure comprised of $10,\!000$ iterations, which required approximately $30$ minutes on the Intel Xeon system. For inference problems involving an exponential kernel, PMMH estimation may be preferred when only one sample path is to be analysed. For settings where an amortised procedure is beneficial, the NN estimator is preferred as estimation accuracy is comparable to PMMH at a lower cost per estimate.

\subsection{Number of Lags}\label{ssec:nlags}
To fit a Hawkes process model with a non-exponential kernel, one must choose the number of lags, $p$, to obtain the NBAR estimates. As demonstrated in Section~\ref{ssec:nn_simstudy_pmmh}, using only very few lags can produce accurate results. However, when the mean and variance of the excitation kernel are large relative to the interval width, the self-excitation effects will typically be realised a number of intervals after a given event. Therefore, the performance of the NN estimator generally improves as $p$ increases. Given that the NBAR($p$) estimates are produced in linear time, increasing the number of lags does not greatly impact the overall time of the estimation procedure. However, large $p$ values may demand an increased number of nodes and training samples to handle the dimension of the summary statistic.

To illustrate the impact of varying $p$ on the resulting NN estimator, Table~\ref{tab:num_p} presents the NN estimation of a Hawkes process with $\mathrm{Gamma}(1.5, 1.0)$ excitation kernel and interval width $\Delta =  0.1$. A larger value of $p$ than that used in Section~\ref{ssec:nn_simstudy_pmmh} will be needed for the best performance, as the $95\%$ quantile of the offspring distribution is now approximately $39\Delta$.

\begin{table}[ht]
    \centering
    \caption{NN estimates with Gamma kernel, $T =  1,\!000$, $\Delta =  0.1$, and varying number of lags, $p$. NN model: two hidden layers with 128 and 64 nodes, respectively.}
    \begin{tabular}{lccccccccccc}\toprule
     & & $\nu$ &$\eta$ & $\alpha$ &$\beta$ & & &$\nu$ &$\eta$ & $\alpha$ &$\beta$ \\
     \midrule
     & & $2.0$ &$0.6$ & $1.5$ &$1.0$ & & &$2.0$ &$0.6$ & $1.5$ &$1.0$ \\ 
    \midrule

\multirow{3}{*}{$p\, =\, 6$} & Est & 1.994 & 0.594 & 1.777 & 0.898 &  \multirow{3}{*}{$p\, =\, 12$} & Est & 2.051 & 0.587 & 1.677 & 0.925 \\
& SE & 0.223 & 0.044 & 0.484 & 0.364 && SE  & 0.207 & 0.041 & 0.437 & 0.315 \\
& CP & 0.958 & 0.948 & 0.986 & 0.936 && CP & 0.942 & 0.950 & 0.914 & 0.898 \\


\midrule

\multirow{3}{*}{$p\, =\, 24$} 
& Est & 2.050 & 0.593 & 1.450 & 1.041 
& \multirow{3}{*}{$p\, =\, 48$}
& Est & 2.064 & 0.591 & 1.528 & 1.046\\
& SE  & 0.218 & 0.040 & 0.304 & 0.285 && SE& 0.207 & 0.040 & 0.291 & 0.290 \\
& CP & 0.962 & 0.950 & 0.976 & 0.984 && CP & 0.968 & 0.966 & 0.968 & 0.952 \\

    \bottomrule
\end{tabular}
    
    \label{tab:num_p}
\end{table}

The number of lags has minimal impact on the estimation of $\nu$ and $\eta$, as these estimates are primarily driven by the imputation component of the summary statistic. With only $p =  3$, the NN estimation of $\alpha$ is noticeably biased. Much of this bias is removed with only a modest increase to $p =  6$, then again increasing to $p =  12$, with a drop in standard error also evident with both moves. Increasing to the larger values of $p =  24$ and then $p =  48$ eventually removes almost all empirical bias from the estimator, with the standard error stabilising.

In light of the results above, some practical recommendations for selecting an adequate number of lags are as follows. Firstly, one can trial different values of $p$, ceasing to increase once estimates stabilise. This requires the training of multiple NNs and is therefore more time consuming. As a quick alternative, one can inspect the estimated $\mathrm{NBAR}(p)$ coefficients and choose a value of $p$ that captures lags with a magnitude that meaningfully differs from zero. 

\subsection{Time-Varying Baseline}
\label{ssec:tvb_simstudy}
In Section~\ref{ssec:timevary}, we proposed a method for obtaining a NN estimate when the underlying Hawkes process is specified with a time-varying background rate. In this section, the method is illustrated using a background rate function of the form 
\begin{align*}
    \nu(t) \ &= \ \nu_1 \ + \ \nu_2\sin\big(2\pi t/100\big).
\end{align*}
This represents an undulating background rate, which is relevant for processes that exhibit seasonal fluctuations in events with known periodicity. For this example, the background parameters $\tht_\nu$ must satisfy the constraints $\nu_1 > 0$ and $\nu_2 < \vert \nu_1\vert$ to ensure that $\nu(t) > 0$ for all $t\in \bb R_+$. The prior over $\tht_\nu$ is taken to be $\nu_1 \sim \mathrm{ISN}(4,\, 3)$, with $\nu_2\mid \nu_1 \sim U(-\nu_1,\, \nu_1).$
The process has a $\mathrm{Gamma}(\alpha, \beta)$ excitation kernel, and we choose the number of lags to be $p_\Delta =  1/\Delta$, as in Section~\ref{ssec:gamma_kern}. The true parameter and associated estimates are displayed in Table~\ref{tab:sin_kernel}. The parameters of the background rate are accurately estimated, as well as those of the offspring kernel, with well calibrated quantile estimates.

\begin{table}[ht]
    \centering
    \caption{NN estimates of with time-varying background rate, Gamma kernel and $T =  1,\!000$. NN model: two hidden layers with 128 and 64 nodes, respectively.}
    \begin{tabular}{lcccccc}\toprule
     & & $\nu_1$ & $\nu_2$ &$\eta$ & $\alpha$ &$\beta$ \\
     \midrule
     & & $5.0$ & $3.0$ &$0.6$ & $1.5$ &$0.25$ \\ 
    \midrule
\multirow{3}{*}{$\Delta\, =\, 0.1$} 
& Est & 4.999 & 3.012 & 0.595 & 1.503 & 0.256 \\
& SE  & 0.267 & 0.223 & 0.021 & 0.148 & 0.036 \\
& CP & 0.936 & 0.954 & 0.934 & 0.958 & 0.910 \\
\midrule
\multirow{3}{*}{$\Delta\, =\, 0.5$} 
& Est & 4.994 & 2.992 & 0.599 & 1.459 & 0.274 \\
& SE  & 0.238 & 0.227 & 0.020 & 0.265 & 0.063 \\
& CP & 0.959 & 0.942 & 0.966 & 0.972 & 0.964 \\
    \bottomrule
\end{tabular}
    
    \label{tab:sin_kernel}
\end{table}

We now repeat the experiment, but specify the Hawkes process with an $\mathrm{Exp}(\beta)$ excitation kernel. As discussed in Section~\ref{sec:sumstat}, in this case, the imputation estimate only is the summary statistic. Table~\ref{tab:sin_kernel_exp} displays the results of this simulation experiment. The parameter $\beta$ is well estimated in this case, illustrating that the $\mathrm{NBAR}$ estimates are not required, though some more significant bias emerges for the extreme aggregation level of $\Delta = 5.0$, which corresponds to an average of $62.5$ events per censoring interval.

\begin{table}[ht]
    \centering
    \caption{NN estimates with time-varying background rate, Exponential kernel and $T =  1,\!000$. NN model: two hidden layers with 64 and 32 nodes, respectively.}
    \begin{tabular}{lccccccccccc}\toprule
     & & $\nu_1$ & $\nu_2$ &$\eta$ & $\beta$ & & & $\nu_1$ & $\nu_2$ &$\eta$ & $\beta$ \\
     \midrule
     & & $5.0$ & $3.0$ &$0.6$ & $0.25$ & & & $5.0$ & $3.0$ &$0.6$ & $0.25$ \\ 
    \midrule
\multirow{3}{*}{$\Delta\, =\, 0.1$} & Est &  4.986 & 2.953 & 0.601 & 0.252 & \multirow{3}{*}{$\Delta\, =\, 0.5$} & Est & 5.102 & 3.063 & 0.595 & 0.250\\
& SE & 0.219 & 0.193 & 0.018 & 0.014 &  & SE & 0.247 & 0.195 & 0.020 & 0.020\\
& CP & 0.936 & 0.976 & 0.968 & 0.980 && CP & 0.942 & 0.988 & 0.948 & 0.990 \\

%
\midrule
\multirow{3}{*}{$\Delta\, =\, 1.0$} & Est & 5.075 & 2.994 & 0.598 & 0.249 & \multirow{3}{*}{$\Delta\, =\, 5.0$} & Est & 4.987 & 2.894 & 0.603 & 0.341\\
& SE & 0.276 & 0.233 & 0.023 & 0.037 & & SE & 0.346 & 0.238 & 0.029 & 0.115\\
& CP & 0.976 & 0.978 & 0.970 & 0.898 && CP & 0.970 & 0.970 & 0.966 & 0.978\\
    \bottomrule

\end{tabular}
    
    \label{tab:sin_kernel_exp}
\end{table}

\section{Applications: Infectious Diseases}\label{sec:app_infec_dis}
Infectious diseases in a fixed geographic area are an ideal candidate for modelling with the Hawkes process. Typically, an immigrant event represents an individual contracting the disease from an exogenous source or from an individual in a different region, whereas offspring events represent the transmission between individuals within the region. Due to the difficulties associated with identifying precise infection times for each individual case, as well as the administrative costs of disease notification systems, infectious diseases are often reported as aggregated weekly counts.

Our first application is to replicate the analysis performed by \cite{cheyssonSpectralEstimationHawkes2022} and \cite{chenEstimatingHawkesProcess2025} on weekly measles counts across Tokyo, Japan, using our NN estimator. The results are consistent with the observed data and the PMMH estimates. Since many infectious diseases exhibit seasonal fluctuations in infection rates due to temperature changes, we then use the NN method to estimate two separate Hawkes process models of Salmonella infections across the state of New South Wales (NSW), Australia, using time-varying background rates. This is a more sound approach to infectious disease modelling, with the NN estimator able to accurately capture the underlying seasonality.

\subsection{Measles in Tokyo}
Weekly counts of measles cases in the greater Tokyo area of Japan were used by both \cite{cheyssonSpectralEstimationHawkes2022} and \cite{chenEstimatingHawkesProcess2025} to demonstrate the efficacy of the PMMH estimator and Whittle estimator, respectively. The PMMH estimator agrees more closely with the observed data than the Whittle estimator, so we use this as the benchmark for comparison. The dataset includes $392$ observations, from the $10$\thh of August, 2012, to the $20$\thh of February 2020. We therefore set $T = 392$ and $\Delta = 1.0$. \cite{chenEstimatingHawkesProcess2025} fit a Hawkes process with an exponential kernel, with additional estimates using a Gamma and Weibull kernel showing very little difference from the exponential estimates. We therefore fit an exponential Hawkes process using the proposed methodology. The NN is trained on $M = 50,\!000$ training samples, with $\nu^{(1:M)}$ and $\beta^{(1:M)}$ obtained via ISN sampling with $\mu_\beta = 4$ and $\sigma_\beta = 2.5$. Table~\ref{tab:tokmeas} shows the NN and PMMH estimates, with their associated quantile estimates (denoted by $\bs{\hth}_{q}$). The NN and PMMH estimates closely agree.

\begin{table}[ht]
    \centering
    \caption{NN and PMMH estimates for weekly measles cases in Tokyo, 10/08/2012 -- 20/02/2020. NN Model: two hidden layers with 64 and 32 nodes, respectively. The NN estimator produces similar results to PMMH.}
    \begin{tabular}{lcccclcccc}\toprule
    & & $\nu$ & $\eta$ & $\beta$\ &&& $\nu$ & $\eta$ & $\beta$\\
    \midrule
    \multirow{3}{*}{NN} & Est & 0.170 & 0.746 & 1.205 & \multirow{3}{*}{PMMH} & Est & 0.170 & 0.745 & 1.181\\
    &$\bs{\hth}_{0.025}$ & $0.107$ & $0.622$ & $0.769$ && $\bs{\hth}_{0.025}$ & $0.138$ & $0.618$ & $0.720$\\
    &$\bs{\hth}_{0.975}$ & $0.233$ & $0.877$ & $1.760$ && $\bs{\hth}_{0.975}$ & $0.202$ & $0.872$ & $1.642$\\
    \bottomrule
\end{tabular}
    
    \label{tab:tokmeas}
\end{table}

\subsection{Salmonella in New South Wales}
Salmonella infection is a type of bacterial illness contracted by humans due to the presence of the Salmonella bacteria in food that has been poorly stored or prepared. Humans who have contracted the infection can spread it to nearby individuals through skin or surface contact, shared food, or shared utensils \citep{SAHealth}. This makes the spread of Salmonella an ideal candidate for modelling using the Hawkes process. An incubation period of typically 12 to 36 hours precedes the infectious period of the disease, which is highly variable and can last several days to multiple weeks \citep{NSWHealth}. An important feature of Salmonella infection is that the number of events increases significantly through the summer months, as higher temperatures provide ideal conditions for the bacteria to grow in unrefrigerated meat \citep{cdcSalmonellaInfection2024}. We therefore require a non-linear background rate to adequately model the process. Seasonal fluctuations in the occurrence rate of infectious diseases are very common, so this analysis highlights the importance of developing estimation techniques that accommodate time-varying background rates in the Hawkes process.

\begin{figure}[ht]
\begin{subfigure}{0.49\textwidth}
\centering
    \includegraphics[width = \textwidth]{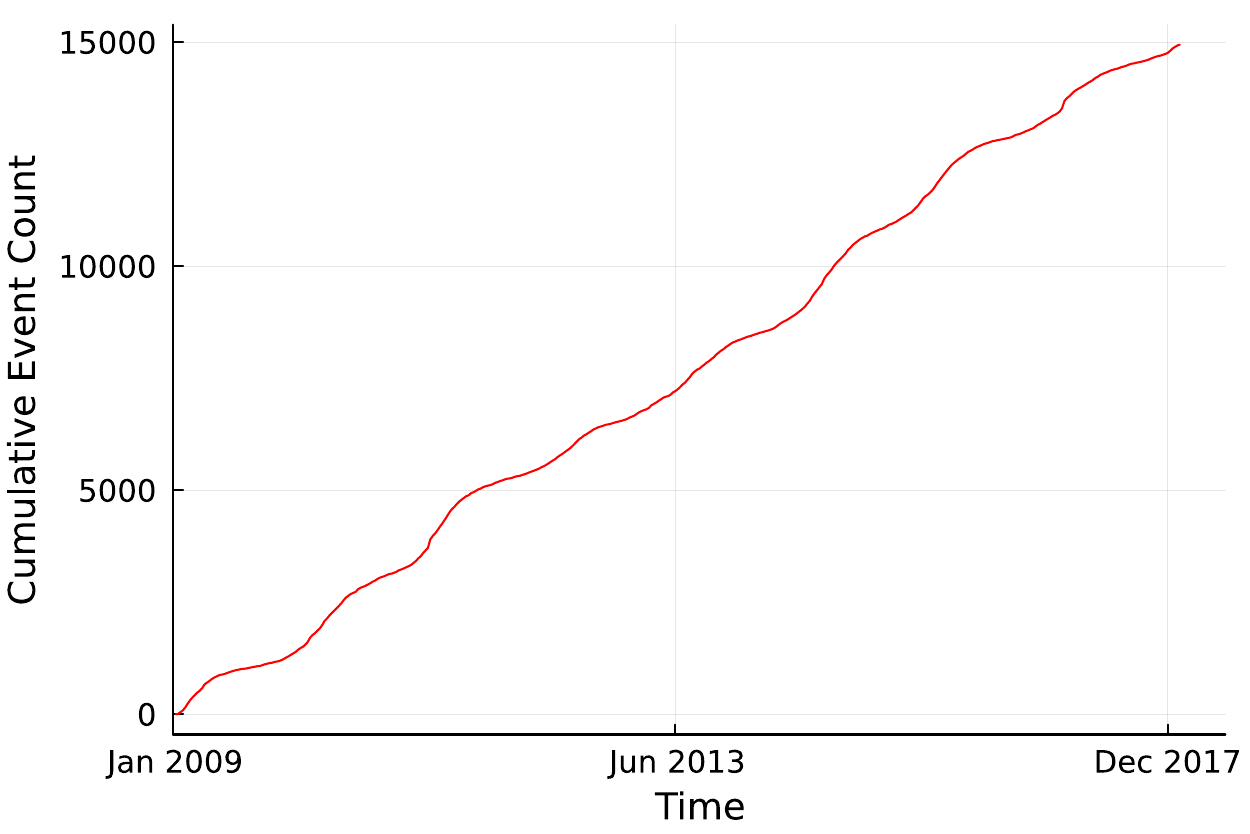}
    \caption{Cumulative event counts}
    \label{fig:cumsal}
\end{subfigure}
\hfill
\begin{subfigure}{0.49\textwidth}
\centering
    \includegraphics[width = \textwidth]{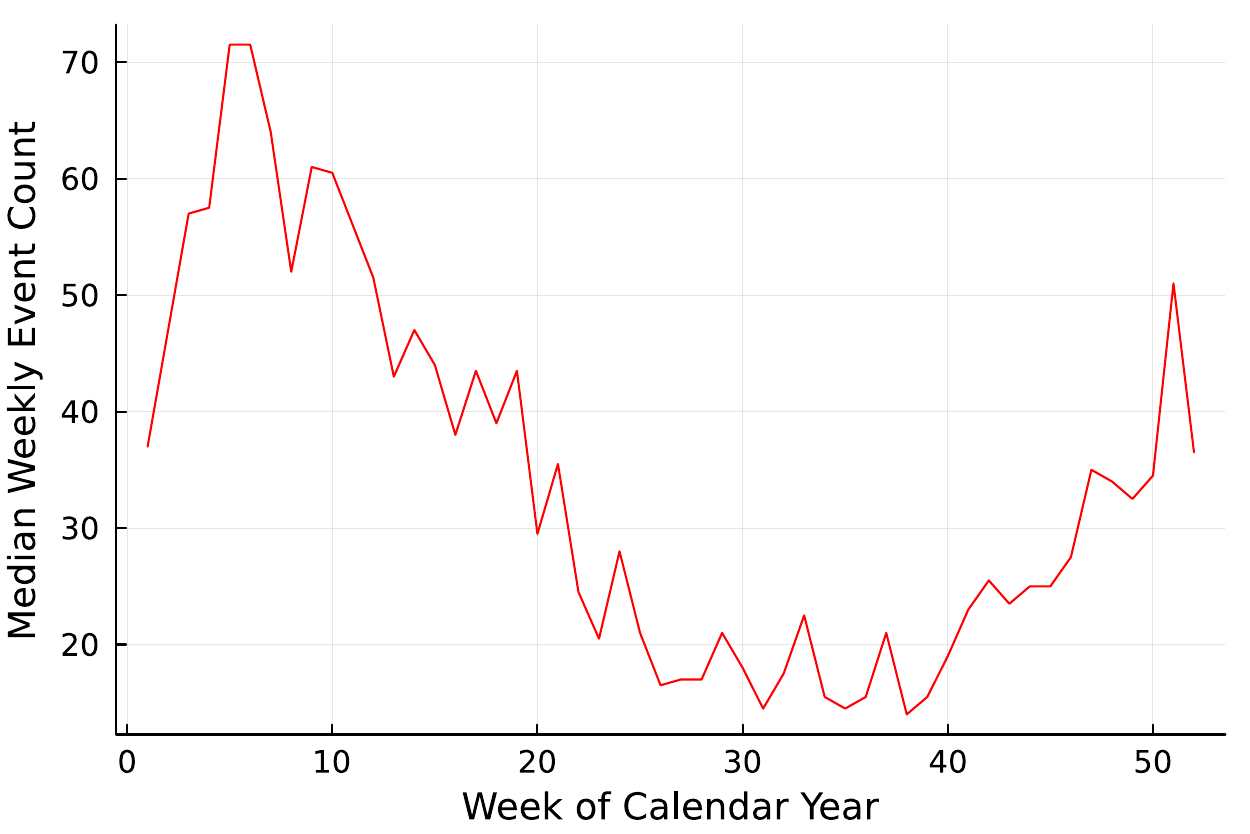}
    \caption{Median weekly event counts}
    \label{fig:wkave}
\end{subfigure}
    \caption{Salmonella Typhimurium Cases in NSW, Jan 2009 -- Dec 2017}
    \label{fig:sal1}
\end{figure}

The National Notifiable Disease Surveillance System (NNDSS) has published weekly Salmonella infection counts across New South Wales (NSW) from 01 Jan 2009 to 31 Dec 2024 \citep{nnds2024}. The strain of the infection for each individual case is also identified in the data set, so we restrict our attention to Salmonella Thyphimurium, as it is most common in NSW. We focus on the period from 01 Jan 2009 to 31 Dec 2017 due to the apparent stability of the underlying dynamics over this time period. Figure~\ref{fig:cumsal} displays the cumulative event counts over the period of interest, alongside Figure~\ref{fig:wkave}, which displays the median weekly event count for each week of the calendar year. The background rate of infection is periodic, as expected from the seasonal changes in Salmonella infection risk. We demonstrate the NN estimation procedure on two possible time-varying background rate functions: a trigonometric function and an order $4$, periodic B-spline.

\subsubsection{Trigonometric Background}
To handle the periodicity of the event counts, a simple choice of background rate is
\begin{align*}
    \nu^\mathrm{tr}(t) \ &= \ \nu_1 \ + \ \nu_2\sin\big(\pi t/26\big) \ + \ \nu_3\cos\big(\pi t/26\big).
\end{align*}
The linear combination of sine and cosine functions allows for phase estimation of the periodic data. The argument $\pi t/26$ ensures that the background completes one period each calendar year. A single imputation estimate of this model returns $\hat\eta^\mathrm{imp} = 0.745$, which is a preliminary indication of high levels of self-excitation. We elect to use a gamma offspring distribution for our model as it can accommodate settings where transmission typically occurs after some delay. A NN is trained using the procedure for time-varying background rates discussed in Section~\ref{ssec:timevary}, with $p = 20$ lags for the $\mathrm{NBAR}$ summary statistic. Table~\ref{tab:sal_est} presents the resulting NN estimates, denoted by $\hth^\mathrm{tr}$, alongside the quantile estimates ($\bs{\hth}_{q}$).

\begin{table}[ht]
    \centering
        \caption{NN estimates for weekly Salmonella Typhimurium cases, with background rate $\nu^\mathrm{tr}$ and Gamma kernel. NN model: two hidden layers with 256 and 128 nodes, respectively.}
    \begin{tabular}{lcccccc}\toprule
     & $\nu_1$ & $\nu_2$ & $\nu_3$ &$\eta$ & $\alpha$ &$\beta$  \\
     \midrule
     Est & 5.263 & 1.802 & 3.689 & 0.854 & 0.243 & 4.921\\
     $\bs{\hth}_{0.025}$ & 3.761 & 0.999 & 2.740 & 0.780 & 0.188 & 0.513 \\
     $\bs{\hth}_{0.975}$ & 7.491 & 3.056 & 4.783 & 0.910 & 0.514 & 9.239 \\
    \bottomrule
\end{tabular}
    \label{tab:sal_est}
\end{table}

\begin{figure}[h]
\begin{subfigure}{0.49\textwidth}
\centering
    \includegraphics[width = \textwidth]{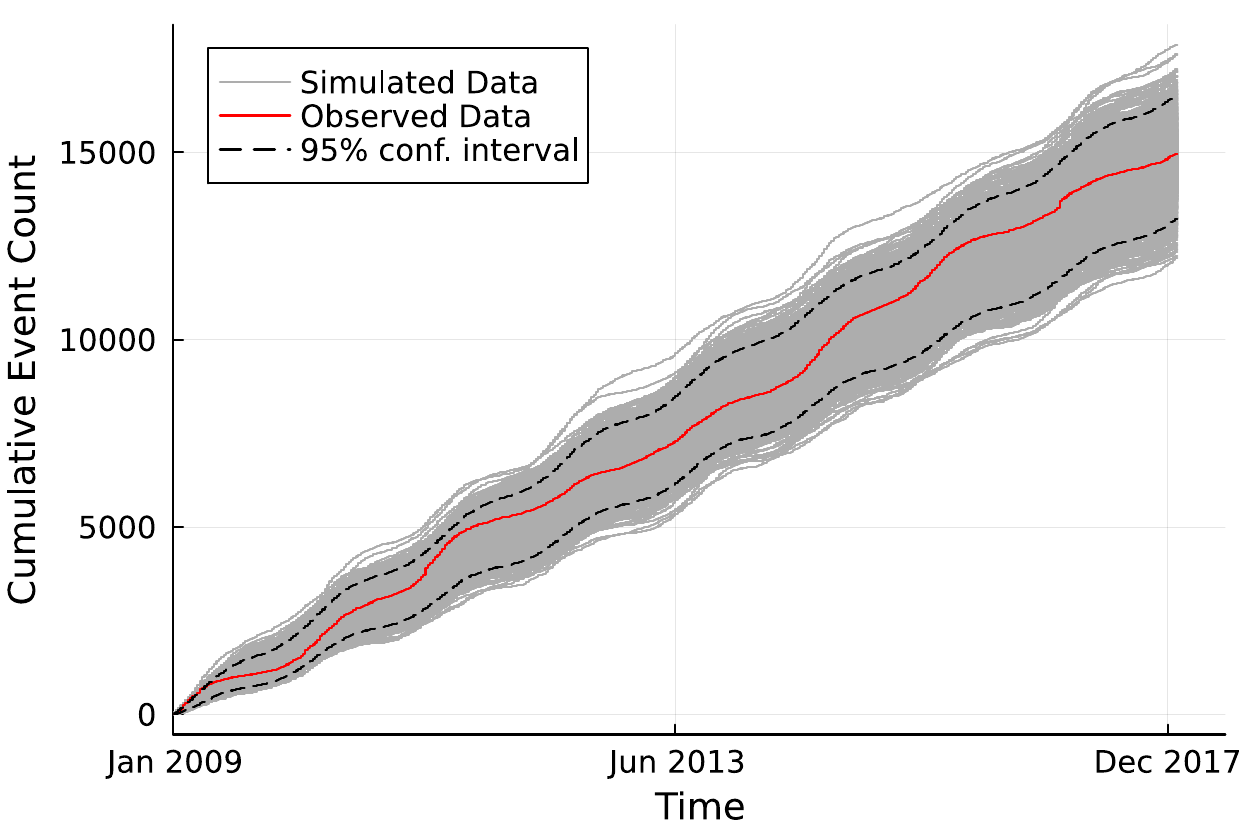}
    \caption{Cumulative event counts}
    \label{fig:cumsal_sim}
\end{subfigure}
\hfill
\begin{subfigure}{0.49\textwidth}
\centering
    \includegraphics[width = \textwidth]{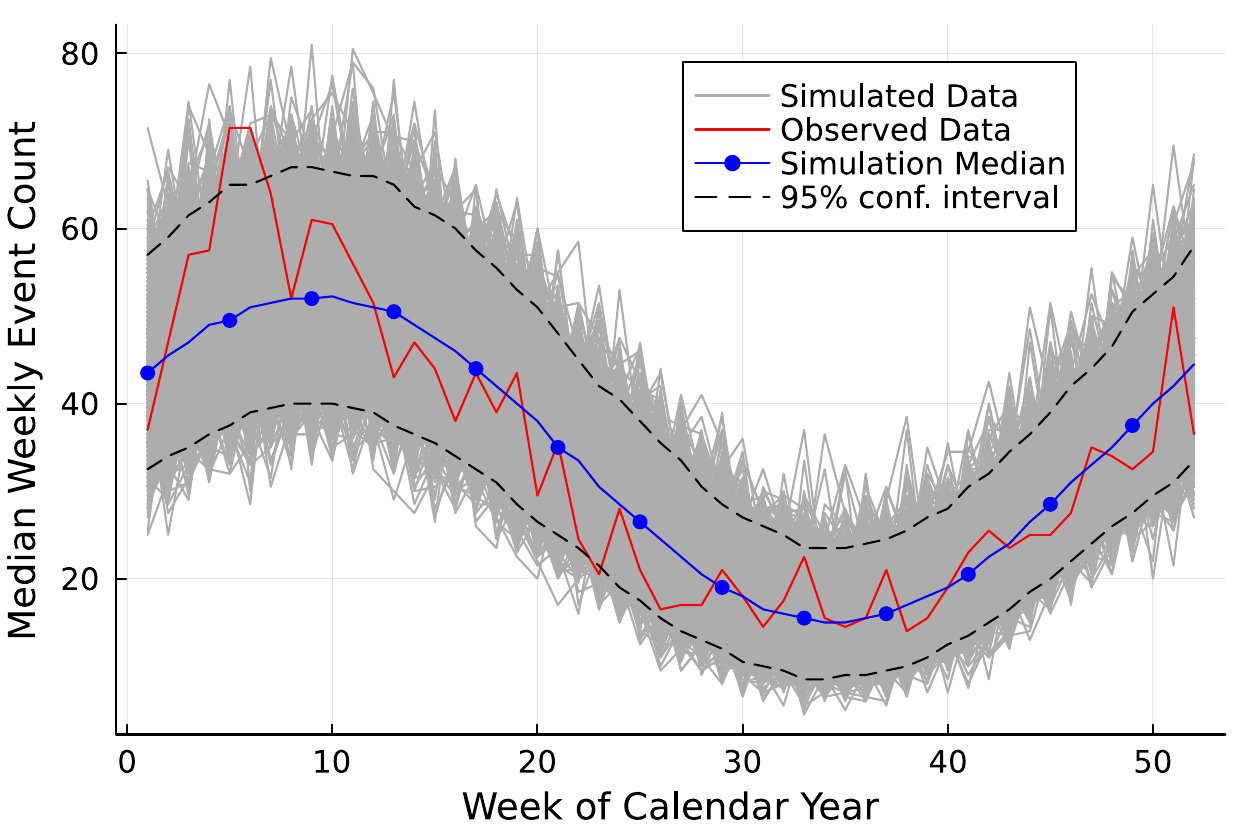}
    \caption{Median weekly event counts}
    \label{fig:wkave_sim}
\end{subfigure}
    \caption{Salmonella Typhimurium Cases in NSW, 2009 - 2017, compared to sample paths simulated from a Hawkes process with a trigonometric background rate function and gamma excitation kernel, and estimated parameter $\hth^\mathrm{tr}$.}
    \label{fig:sal_nn}
\end{figure}

We simulate $1,\!000$ sample paths of the Hawkes process under $\hth^\mathrm{tr}$ and compute the median event counts for each week of the calendar year. Figure~\ref{fig:sal_nn} overlays the observed weekly averages on the simulated paths. The proposed background rate provides a reasonable approximation of the fluctuations in event counts, though it does not fully capture the magnitude of the peak in summer. The observed mean weekly count is $36.196$, with the mean weekly count suggested by our estimator being close to this value, at
\begin{align*}
    \frac{\frac{1}{T}\int_0^{T} \hat\nu^\mathrm{tr}(s)\dd s}{1\, -\, \hat\eta^\mathrm{tr}}\ = \ 35.944.
\end{align*}
An estimate of $0.854 \ (0.780,\, 0.910)$ for the branching ratio suggests very high levels of temporal clustering associated with Salmonella infection cases. Salmonella Typhimurium has an incubation period of $6$ hours to $3$ days, though incubation is typically between $12$ and $36$ hours \citep{NSWHealth}. The infectious period varies by individual, though periods of multiple days to multiple weeks are common \citep{NSWHealth}. The estimates $\hat\alpha^\mathrm{tr}$ and $\hat \beta^\mathrm{tr}$ imply a median time from offspring waiting time of 1.38 days, which is slightly lower than what is implied by the reported incubation and infectious periods. However, the long tail of the estimated offspring kernel supports the possibility of a few cases resulting in a prolonged infectious period. 

\subsubsection{Spline Background}
The trigonometric background rate is simple and computationally efficient to implement, though it underestimates the rate of infection during the peak season. A more flexible model is to define the background rate function using a periodic, order $4$ B-spline, denoted by $S:\bb R_+\to \bb R$. We place five knots over each year, at weeks $\{0,\,2.5,\, 5,\, 38,\, 52\}$, with the first and last knots defining the period. The interior knots at $5$ and $38$ are chosen as they match the empirical minimum and maximum of the median weekly infection count, respectively, with an additional knot at $2.5$ to allow for a rapid increase in background rate during summer. The spline $S$ requires four parameters, $\nu_{1:4}$, to be fully specified. To avoid imposing positivity constraints on the coefficients of the spline function, we define the background rate function to be $\nu^{\mathrm{sp}}(t) \, =\, f\circ S(t)$, where $f$ again denotes the softplus function. This formulation allows for $\nu_{1:4}\subset \bb R^4$, which substantially improves the computational speed of obtaining imputation estimates, whilst retaining the modelling flexibility of a periodic spline. The training sample for each $\nu_i$ is simply drawn from a normal distribution centred around the respective imputation estimate, with relatively large variance. The NN point estimate, $\hth^\mathrm{sp}$, and associated quantile estimates are displayed in Table~\ref{tab:nn_spline}.

\begin{table}[ht]
    \centering
        \caption{NN estimates and quantile estimates for weekly Salmonella Typhimurium cases, with background rate $\nu^\mathrm{sp}$ and gamma kernel. NN model: two hidden layers with 256 and 128 nodes, respectively.}
    \begin{tabular}{lccccccc}\toprule
     & $\nu_1$ & $\nu_2$ & $\nu_3$ & $\nu_4$ &$\eta$ & $\alpha$ &$\beta$  \\
     \midrule
     Est & 6.350 & 13.252 & 2.364 & 0.229 & 0.848 & 0.332 & 3.584\\
     $\bs{\hth}_{0.025}$ & 4.356 & 9.912 & -0.722 & -1.460 & 0.804 & 0.266 & 1.887\\
     $\bs{\hth}_{0.975}$ & 8.462 & 16.563 & 5.259 & 2.009 & 0.888 & 0.416 & 7.426\\
    \bottomrule
\end{tabular}

    \label{tab:nn_spline}
\end{table}

\begin{figure}[ht]
\begin{subfigure}{0.49\textwidth}
\centering
    \includegraphics[width = \textwidth]{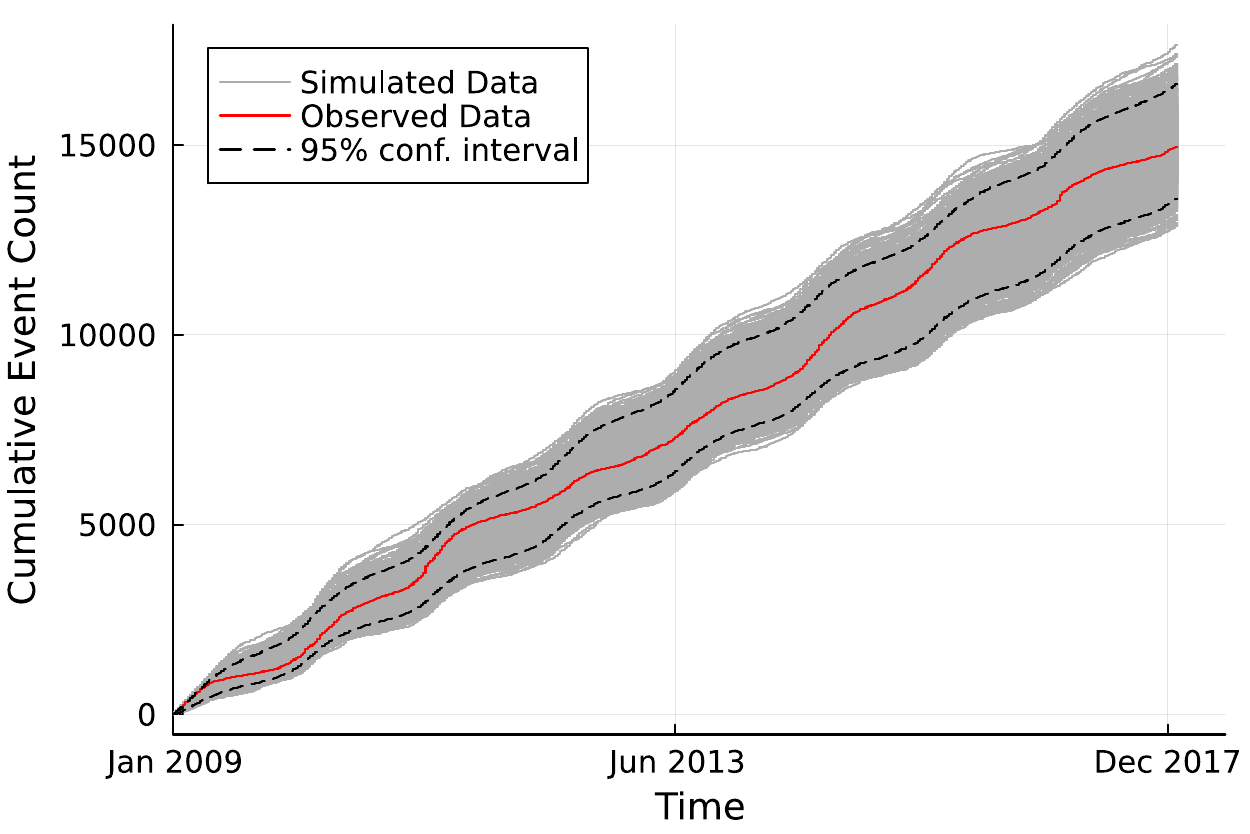}
    \caption{Cumulative event counts}
    \label{fig:cumsal_sim_spline}
\end{subfigure}
\hfill
\begin{subfigure}{0.49\textwidth}
\centering
    \includegraphics[width = \textwidth]{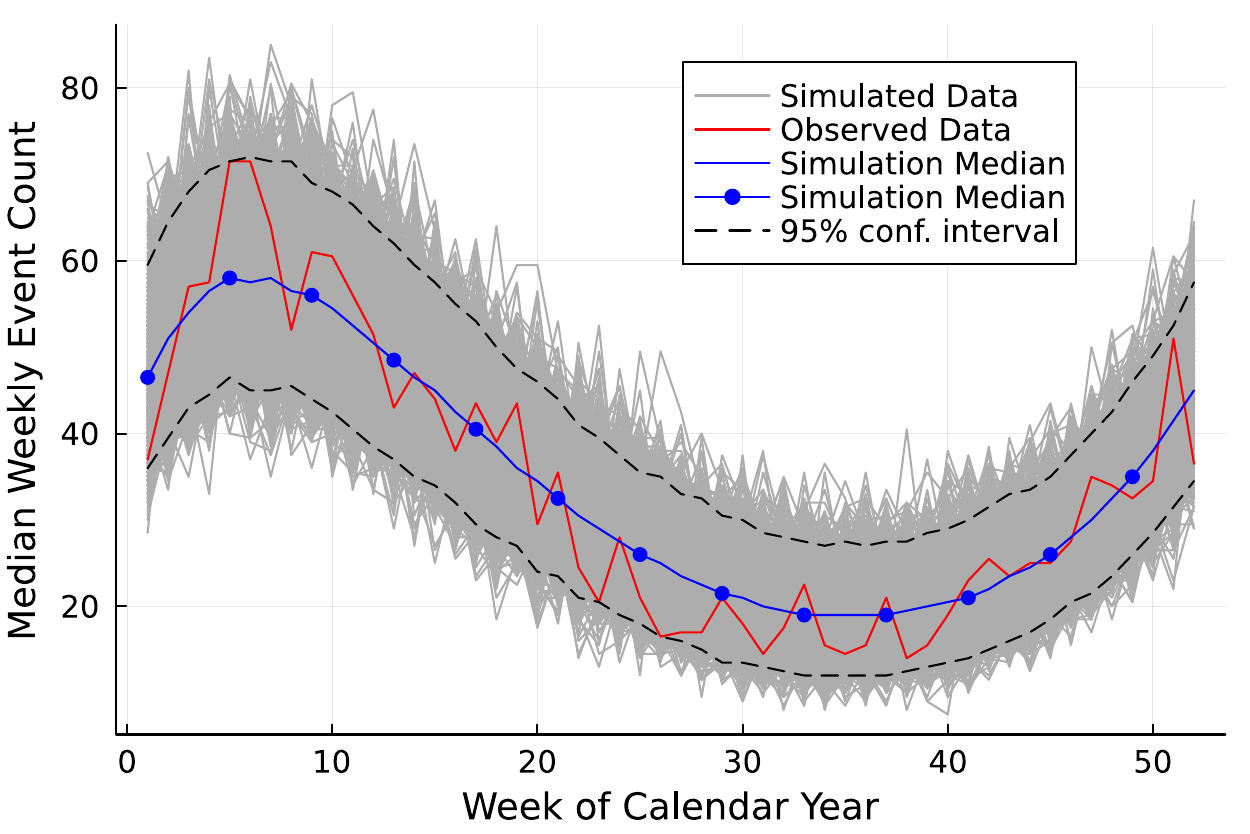}
    \caption{Median weekly event counts}
    \label{fig:wkave_sim_spline}
\end{subfigure}
    \caption{Salmonella Typhimurium Cases in NSW, 2009 - 2017, compared to sample paths simulated from a Hawkes process with a periodic B-spline background rate function and gamma excitation kernel and estimated parameter $\hth^\mathrm{sp}$.}
    \label{fig:sal_nn_spline}
\end{figure}

Figure~\ref{fig:sal_nn_spline} again compares the observed sample paths to those produced from simulations from $\hth^\mathrm{sp}$. The spline clearly better captures the spike in event cases during summer with no weekly medians outside the bootstrap 95\% confidence interval. The estimates $\hat{\eta}^\mathrm{sp}$ and $\hat{\eta}^\mathrm{tr}$ are very close, reinforcing the inference that Salmonella infection exhibits significant temporal clustering in NSW. The mean weekly event rate from the spline estimates is
\begin{align*}
    \frac{\frac{1}{T}\int_0^{T} \hat\nu^\mathrm{sp}(s)\dd s}{1\, -\, \hat\eta^\mathrm{sp}}\ = \ 36.850,
\end{align*}
which is close to the observed value of $36.196$. The kernel parameter estimates $\hat \alpha^\mathrm{sp}$ and $\hat \beta^\mathrm{sp}$ now place the median time between contraction and transmission at 2.37 days, which presents no clear disagreement with known infectious periods for Salmonella \citep{NSWHealth}. Additionally, the estimated posterior quantiles for $\alpha$ and $\beta$ are significantly narrower than when using the trigonometric background, so the use of an adequate background rate model seems to allow for better identification of the excitation kernel.

\section{Discussion}\label{sec:NN_disc}
Our work proposes a likelihood-free approach to parameter estimation for the discretely observed Hawkes process by training a neural network to estimate the parameters from a summary statistic. From our experiments, the neural network estimator has limited empirical bias and standard errors comparable to the benchmark PMMH estimator proposed in \cite{chenEstimatingHawkesProcess2025}. The efficacy of the method relies on our construction of a highly informative summary statistic, consisting of a naive uniform imputation estimate of the parameters, along with an additional negative binomial autoregression of the count data that is used in the non-Markovian setting. Our proposed summary statistic can handle unequal censoring intervals and time-varying baselines, which is an advantage over many extant likelihood-based methods. Furthermore, the NN quantile estimation procedure produces well-calibrated credible intervals, making parametric inference fully amortised.

Our use of a naive imputation estimate as the basis of the summary statistic demonstrates that complex reconstructions of the latent event times \citep{shlomovichUnivariate2022, schneiderEstimationSelfexcitingPoint2023} are not necessary. This reduces the level of expert knowledge required by a statistician in designing useful summary statistics. The notion of using simple imputation to generate a summary statistic is generalisable to other settings where the likelihood is intractable due to incomplete information. Applying our proposed technique to other point processes, such as the renewal Hawkes process \citep{stindlLikelihoodBasedInference2018}, is an interesting avenue for future work. Whether the neural network estimator performs well when extended to the multivariate Hawkes process also remains to be explored. The imputation estimate is still immediately available for use in the summary statistic, though experimentation is required to assess whether a multivariate autoregression allows for the offspring kernel to be adequately estimated.

Finally, we note that the neural networks used to produce the estimates in this work are designed following standard recommendations for neural network regression problems of our given complexity. Many decisions are involved in designing a neural network, including the number and size of the hidden layers, the size of the training sample, the choice of activation functions, and the tuning of many other hyperparameters. Our work illustrates that high-quality estimators can be obtained without extensive tuning, though further improvements in performance and computational efficiency may be possible by tuning various aspects of the neural network architecture.

\addcontentsline{toc}{section}{References}

\bibliography{NN}    

\appendix

\section{NN Estimator Function}\label{app:level_curves}
\begin{figure}[h]
\begin{subfigure}{0.32\textwidth}
\centering
    \includegraphics[width = \textwidth]{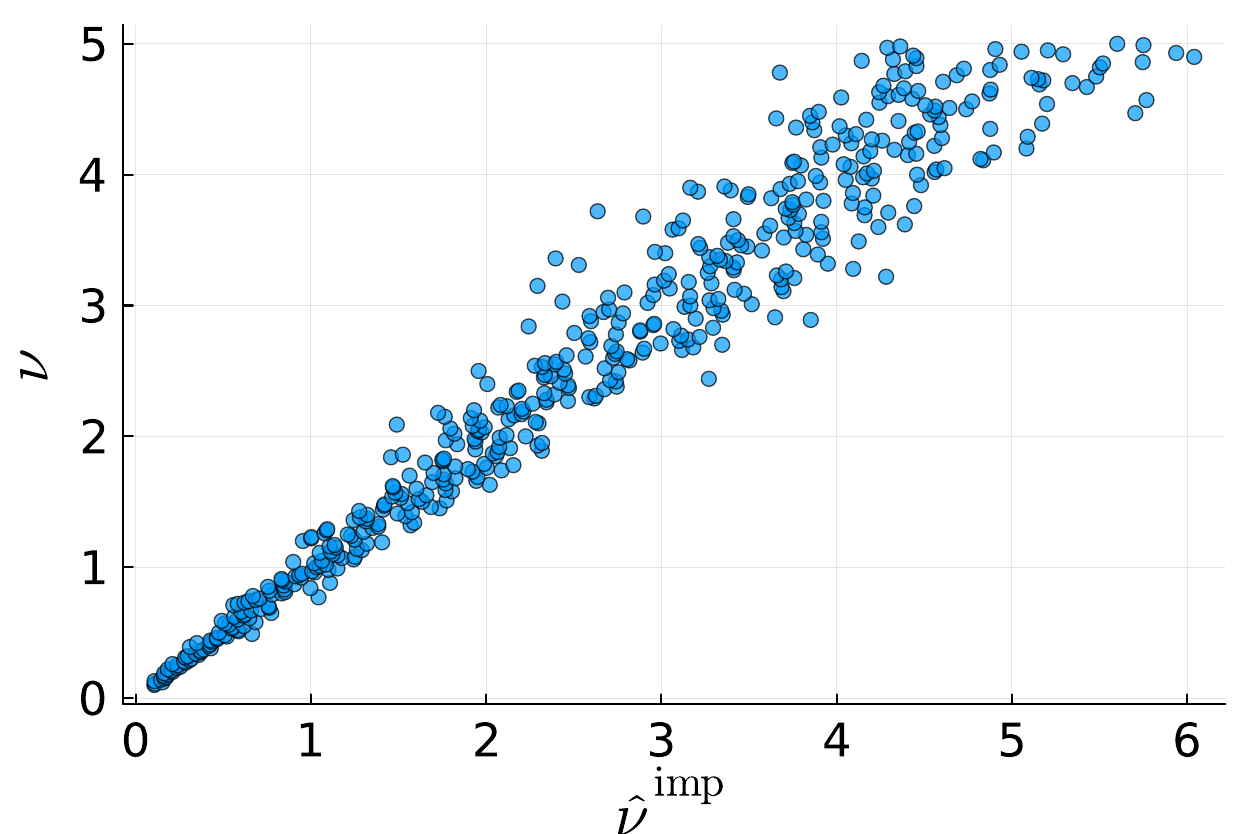}

\end{subfigure}
\hfill
\begin{subfigure}{0.32\textwidth}
    \centering
    \includegraphics[width = \textwidth]{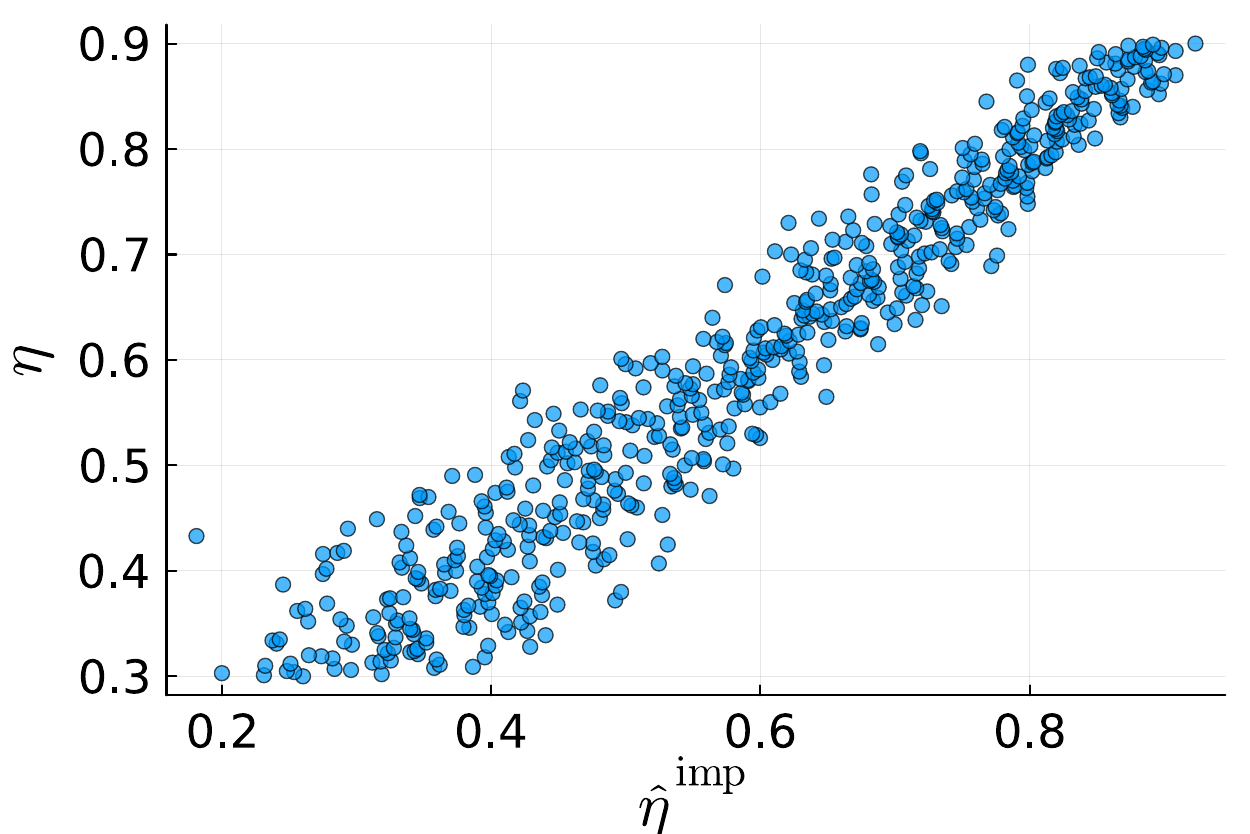}

\end{subfigure}
\hfill
\begin{subfigure}{0.32\textwidth}
    \centering
    \includegraphics[width = \textwidth]{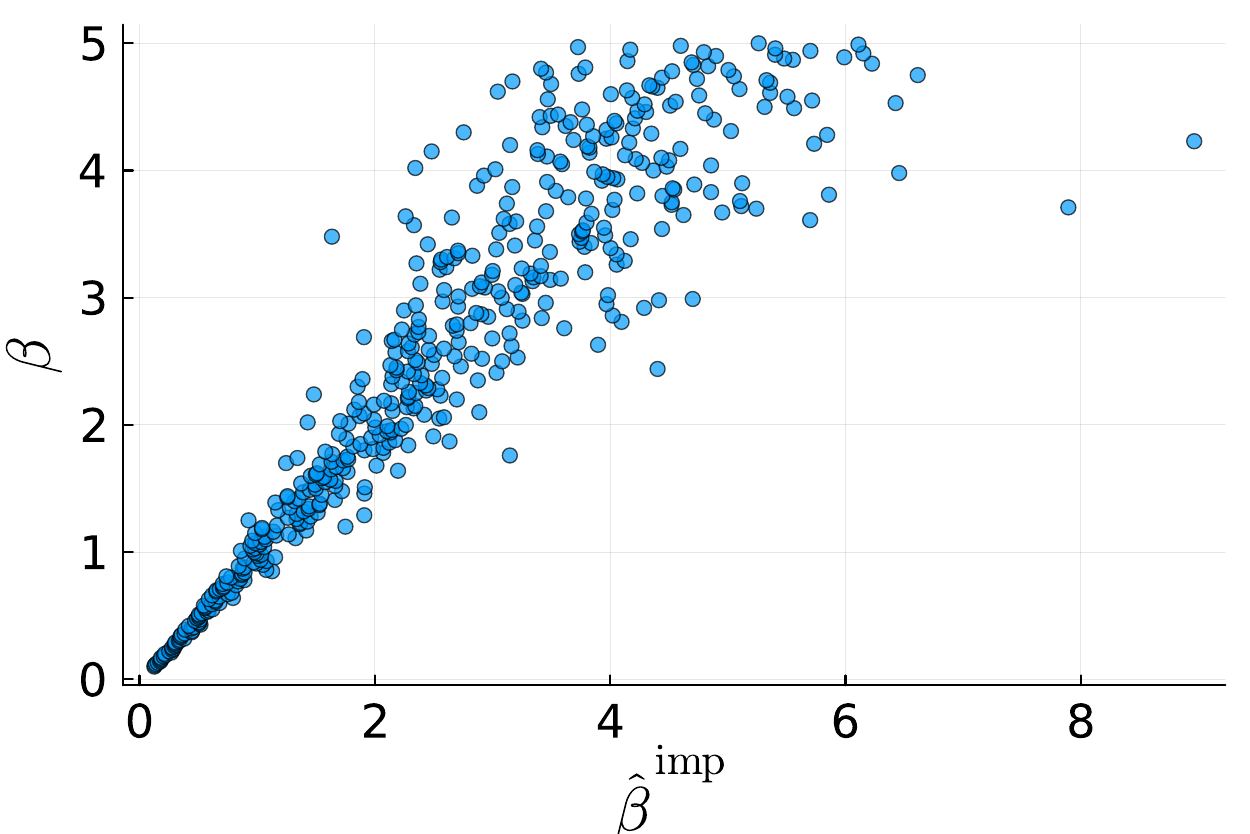}

\end{subfigure}
\\

\begin{subfigure}{0.32\textwidth}
\centering
    \includegraphics[width = \textwidth]{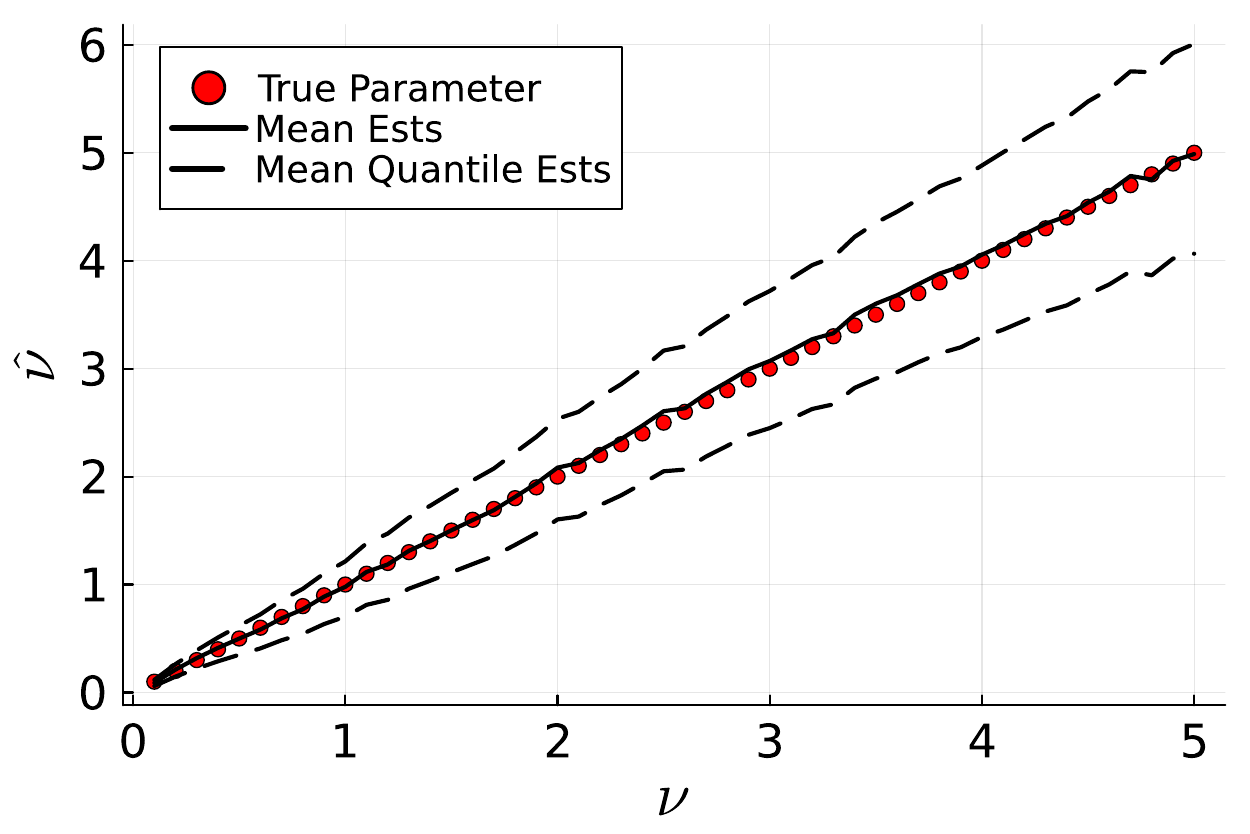}

\end{subfigure}
\hfill
\begin{subfigure}{0.32\textwidth}
    \centering
    \includegraphics[width = \textwidth]{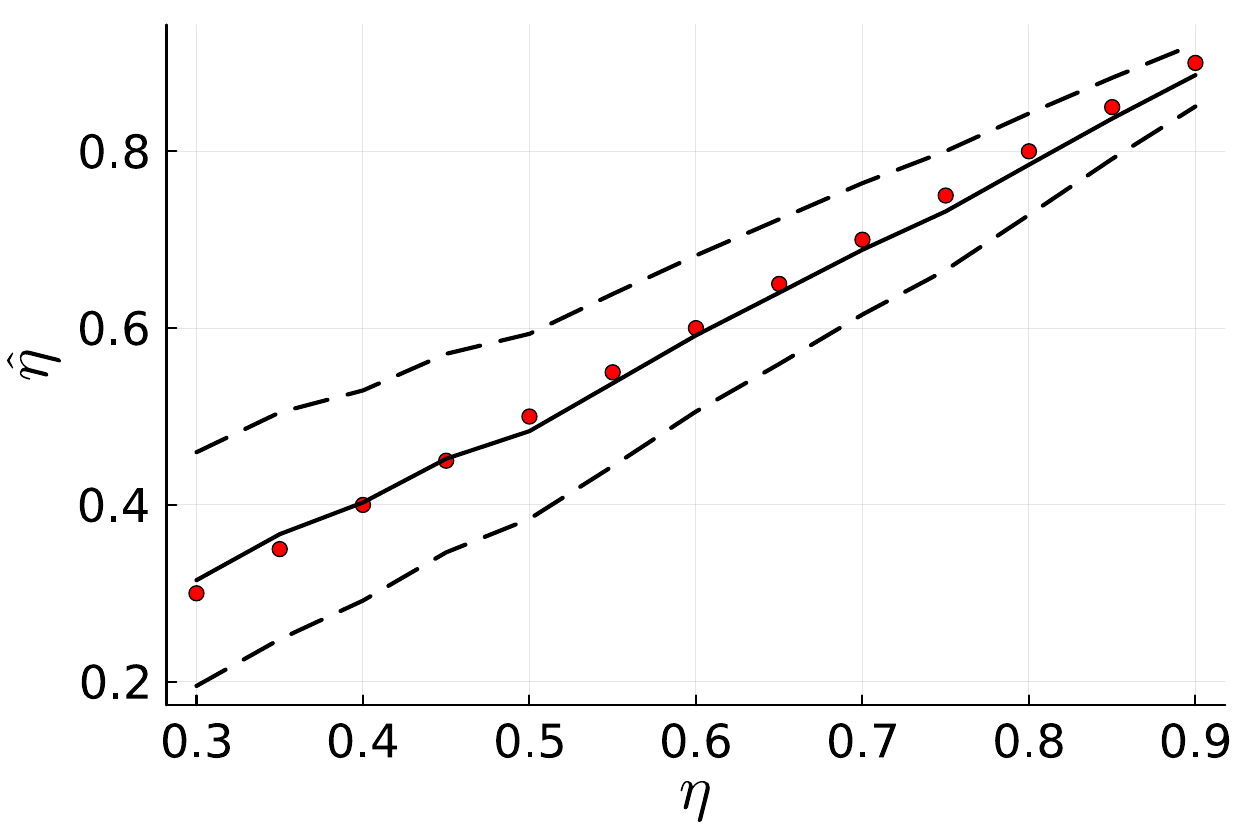}

\end{subfigure}
\hfill
\begin{subfigure}{0.32\textwidth}
    \centering
    \includegraphics[width = \textwidth]{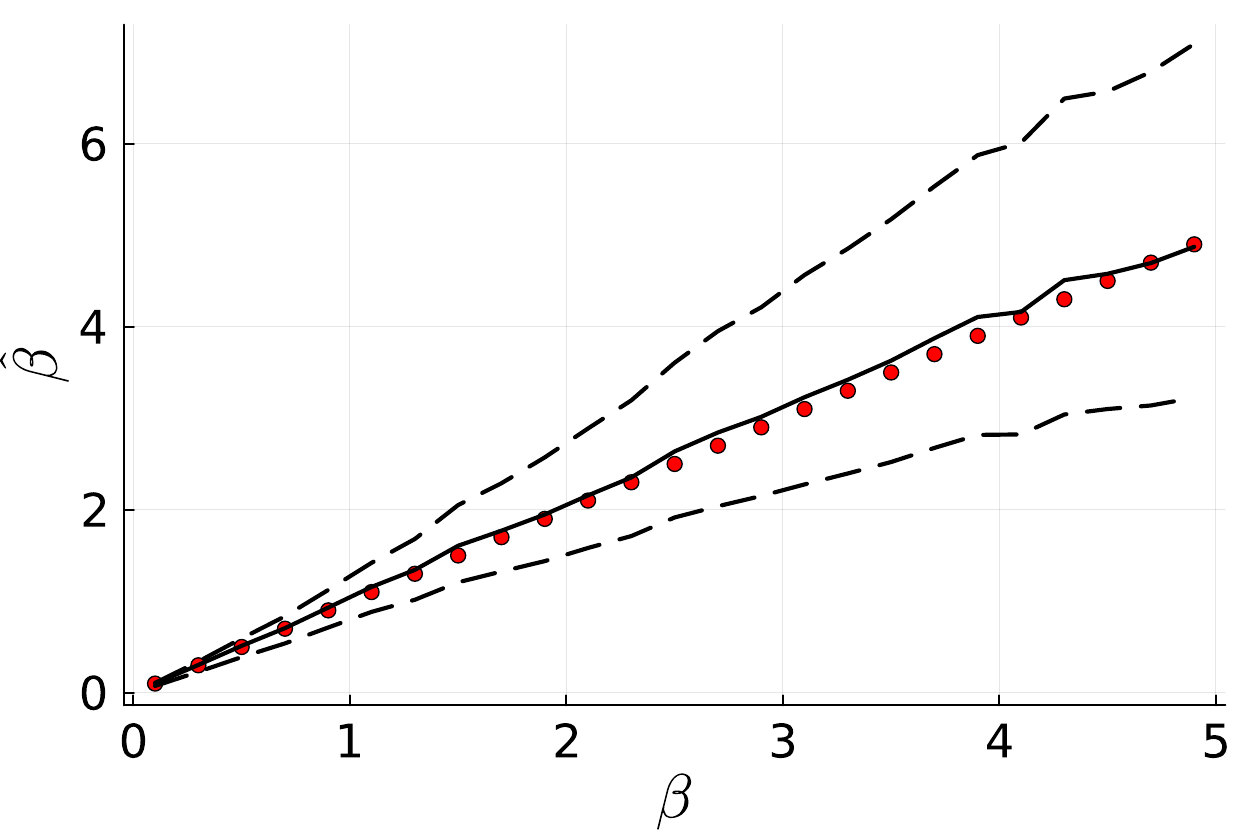}

\end{subfigure}
\\
\begin{subfigure}{0.32\textwidth}
\centering
    \includegraphics[width = \textwidth]{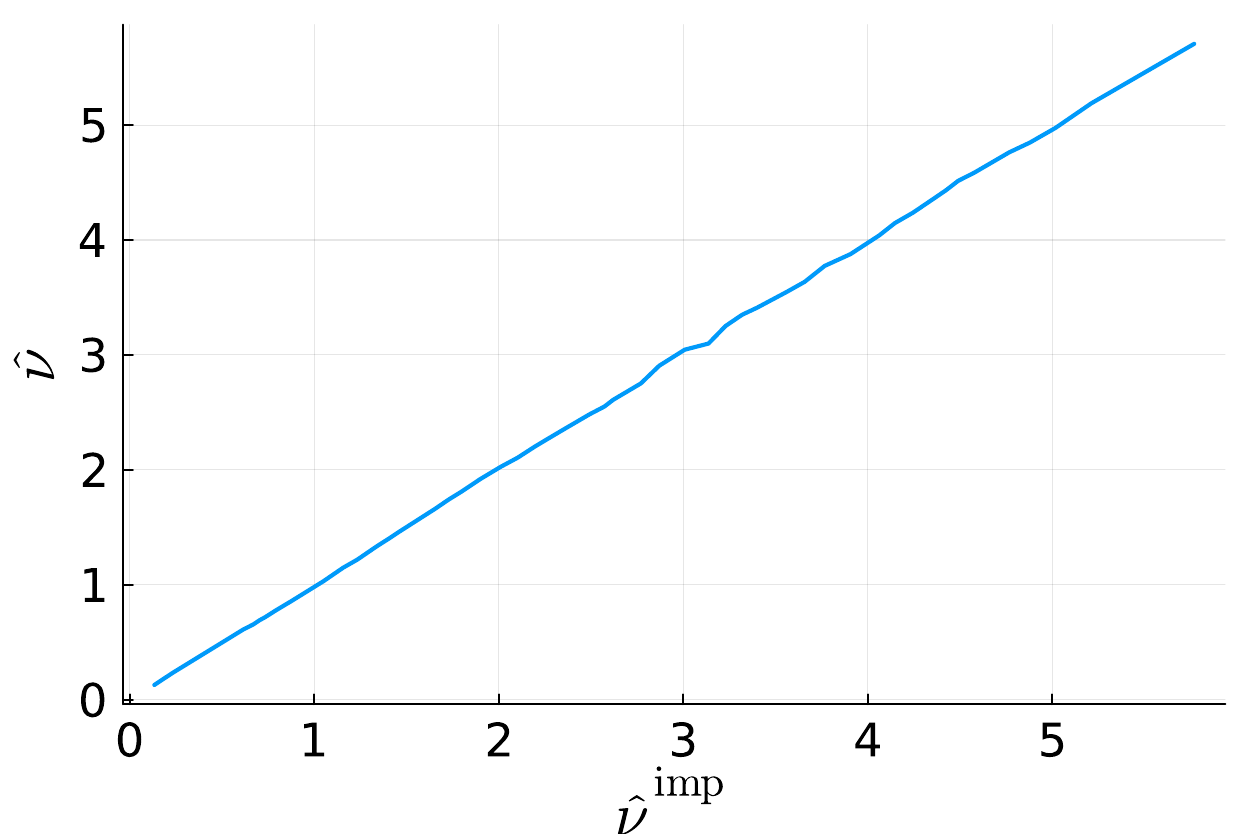}
    \caption{Varying $\nu\in [0.1, 5.0]$}
\end{subfigure}
\hfill
\begin{subfigure}{0.32\textwidth}
    \centering
    \includegraphics[width = \textwidth]{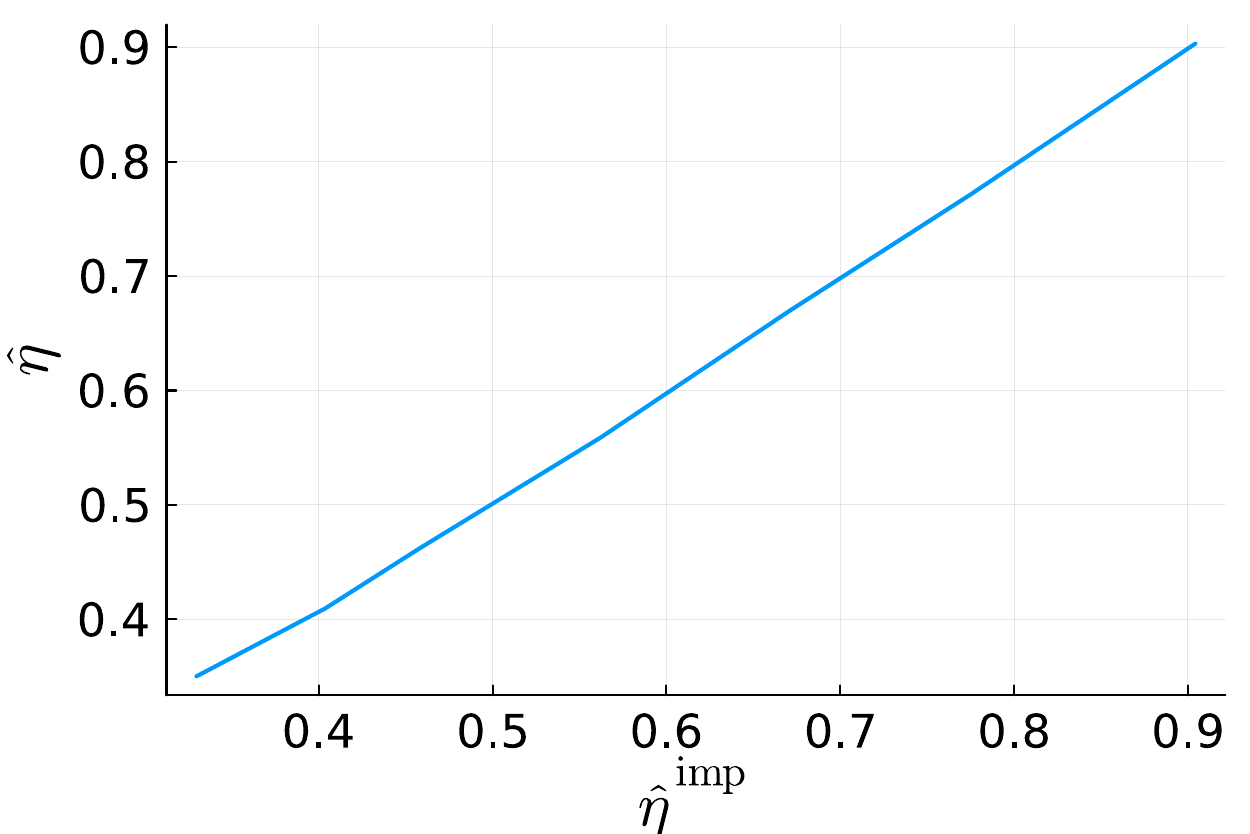}
    \caption{Varying $\eta\in [0.3, 0.9]$}
\end{subfigure}
\hfill
\begin{subfigure}{0.32\textwidth}
    \centering
    \includegraphics[width = \textwidth]{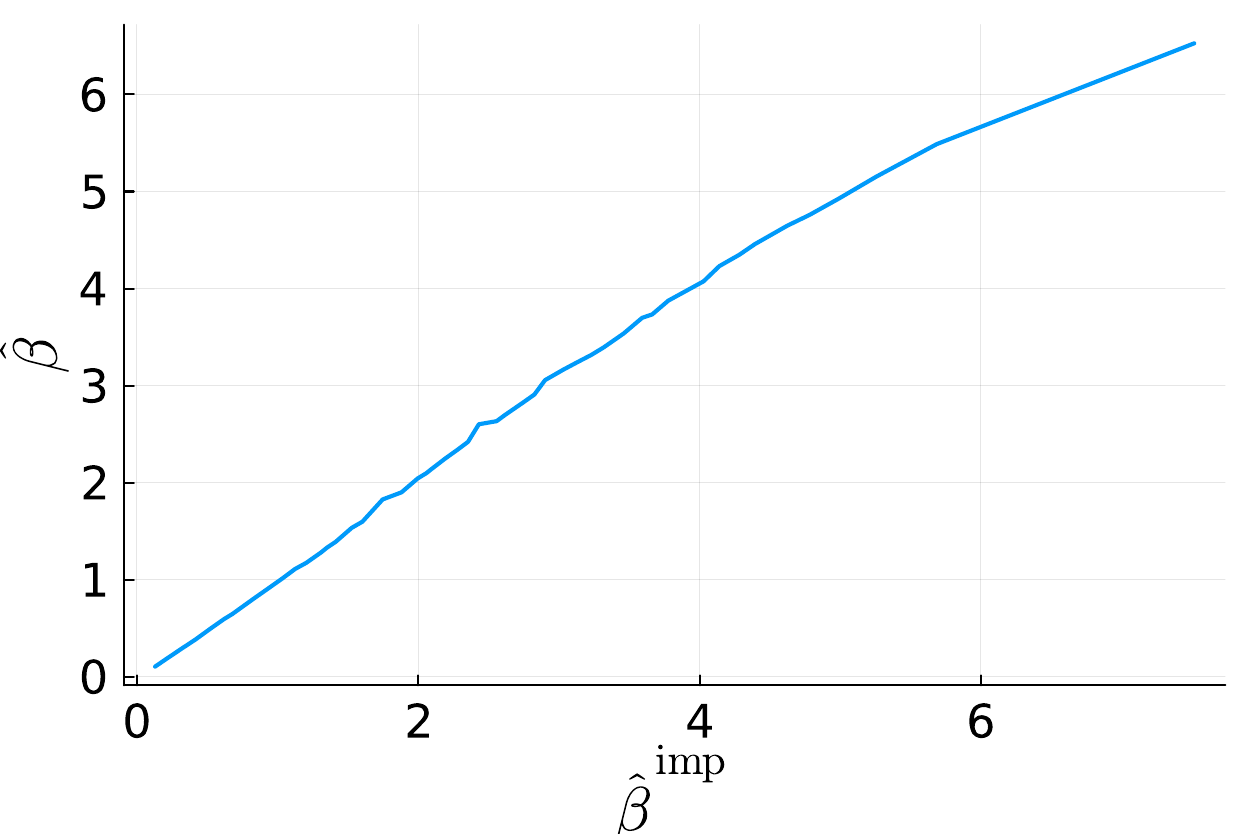}
    \caption{Varying $\beta\in [0.1, 5.0]$}
\end{subfigure}
\hfill

    \caption{Plots of the relationship between the imputation estimate, NN estimate and the true parameter, for $T = 1,\!000$ and $\Delta = 0.1$. Base parameter is $\tht = (2.0, 0.6, 2.0)$ at which the remaining two dimensions are held constant. Top: Plots of the imputation estimates and the true parameters. Middle: Plotting the mean NN estimates against the true parameters. Bottom: Profiles of the trained NN for varied input values.}
    \label{fig:level_curves}
\end{figure}

Consider the discretely observed Hawkes process with censoring time $T$ and aggregation width $\Delta = 0.1$. To provide more clarity on the nature of the trained model in Section 4.1, we now plot profiles of the NN estimator function along each dimension, keeping the remaining two dimensions fixed. The base parameter is again set to be $(\nu, \eta,\beta) = (2.0, 0.6, 2.0)$. Figure~\ref{fig:level_curves} shows the results. For each dimension, the top row plots the imputation estimate against the true parameter, representing the noisy function that the NN is trained to approximate. The middle row plots the true parameter against the mean NN estimate from $100$ simulated sample paths, along with the mean quantile estimates. We see that the trained NN appears to be unbiased for a range of parameter combinations. The bottom row plots the NN estimate against the imputation estimate. One sees from the bottom row that the NN estimator is approximately linear, though there is some curvature along the $\beta$ dimension. More significant curvature is present for larger $\Delta$ values. The NN estimates of the non-varying parameters are horizontal, as expected, and are omitted.

\section{Convergence of summary statistics and the NN estimator}\label{app:sumstats}
\subsection{Background}
Recall that the parameter of interest is $\tht  = (\tht_\nu,\eta, \tht_h)$, where $\tht_\nu$ parametrises the background rate function, and $\tht_h$ parametrises the offspring kernel. In the case of the exponential Hawkes process, our summary statistic is the imputation estimate $\bs s (n_{1:K}) = \hth^\mathrm{imp}$. When working with a non-exponential kernel, the summary statistic is $\bs s(n_{1:K}) = (\hth^\mathrm{imp}, \hat \gamma_{0:p}, \hat \phi)$, with $\hat{\gamma}_{0:p}$ and $\hat\phi$ being the estimated AR coefficients and dispersion parameter, respectively, from a $\mathrm{NBAR}(p)$ model of the count data.

For now, let us restrict our attention to the exponential kernel setting. The imputation estimate is a deterministic function of the data. To be explicit, we write $\hth^\mathrm{imp}(n_{1:K})$. Though the model used to compute $\hth^\mathrm{imp}$ is misspecified, it may still be possible to establish the existence of a limit
\begin{align}\label{eq:imp_lim}
    \lim_{K\to\infty}\hth^\mathrm{imp}(n_{1:K}) \ \eqas \ \tht^*(\tht_0).
\end{align}
The limit $\tht^*(\tht_0)$ will typically not equal the true parameter $\tht_0$, but will depend on $\tht_0$ through its role in the data generating process. This is akin to the classical work of \cite{whiteMaximumLikelihoodEstimation1982} on properties of the MLE under model misspecification. The NN is trained to approximate the mapping $\hth^\mathrm{imp}(n_{1:K}) \mapsto \tht_0$. Taking $K\to\infty$, if the limit exists as in \eqref{eq:imp_lim} and is unique for each $\tht_0$, then the summary statistic is perfectly able to identify the true parameter. This is the sense in which the term \textit{asymptotically sufficient} is used in relation to indirect inference in \cite{drovandiBayesianIndirectInference2015}. 

Proof of the existence of a unique limit of the form in \eqref{eq:imp_lim} is made difficult by the intractable parametric model that specifies the discretely observed Hawkes process. We now present numerical experiments to assess the convergence of the summary statistics, and the rate of convergence of the NN estimator.

\subsection{Convergence of the Summary Statistics}
\begin{figure}[h]
\begin{subfigure}{0.32\textwidth}
\centering
    \includegraphics[width = \textwidth]{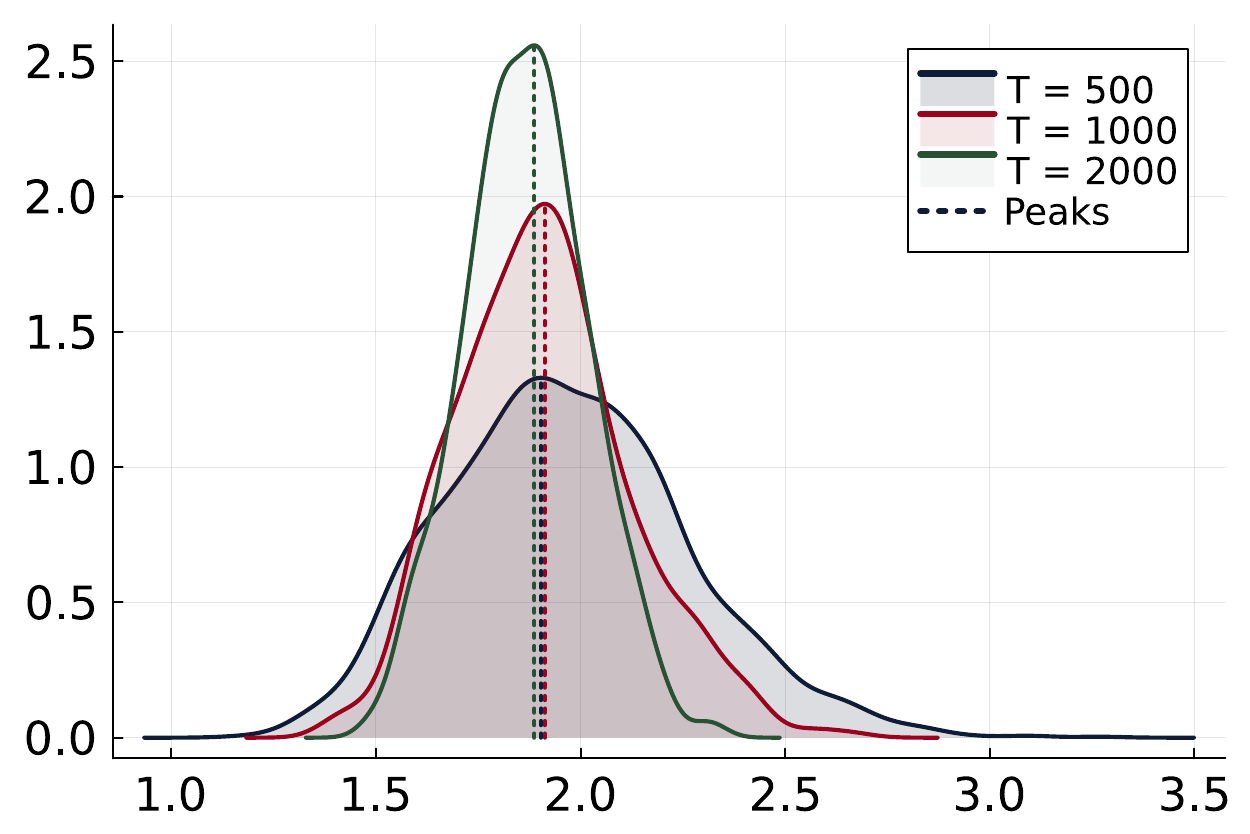}
    \caption{$\hat\nu^\mathrm{imp}$}

\end{subfigure}
\hfill
\begin{subfigure}{0.32\textwidth}
    \centering
    \includegraphics[width = \textwidth]{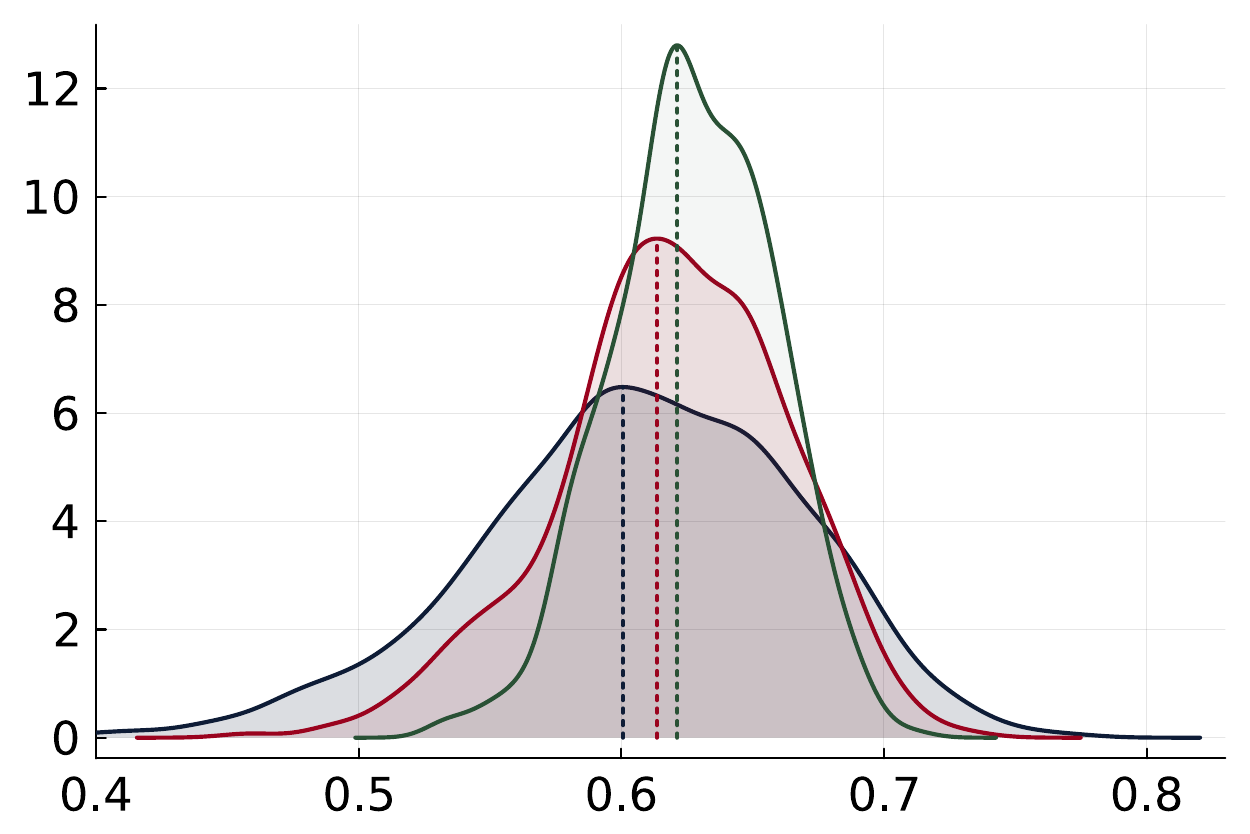}
    \caption{$\hat\eta^\mathrm{imp}$}

\end{subfigure}
\hfill
\begin{subfigure}{0.32\textwidth}
    \centering
    \includegraphics[width = \textwidth]{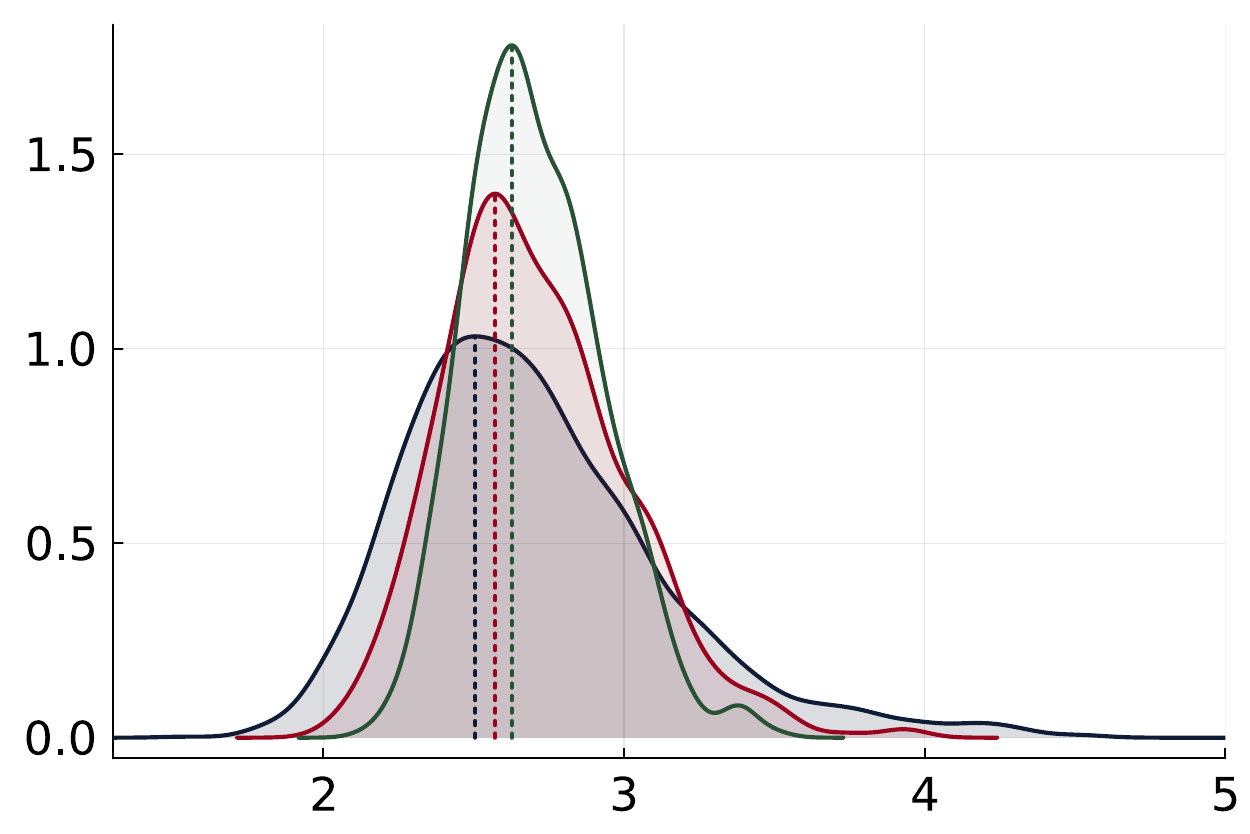}
    \caption{$\hat\beta^\mathrm{imp}$}

\end{subfigure}
\\
\begin{subfigure}{0.32\textwidth}
\centering
    \includegraphics[width = \textwidth]{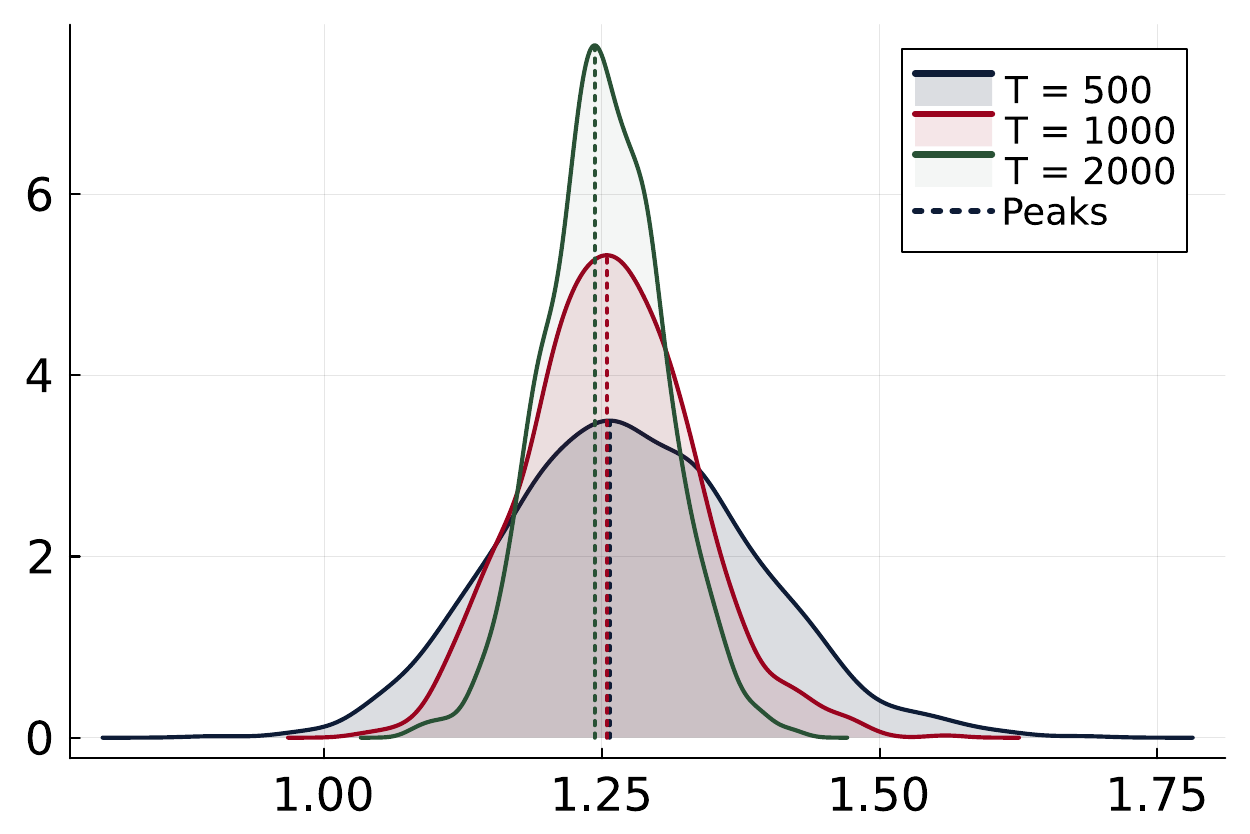}
    \caption{$\hat\gamma_0$}
\end{subfigure}
\hfill
\begin{subfigure}{0.32\textwidth}
    \centering
    \includegraphics[width = \textwidth]{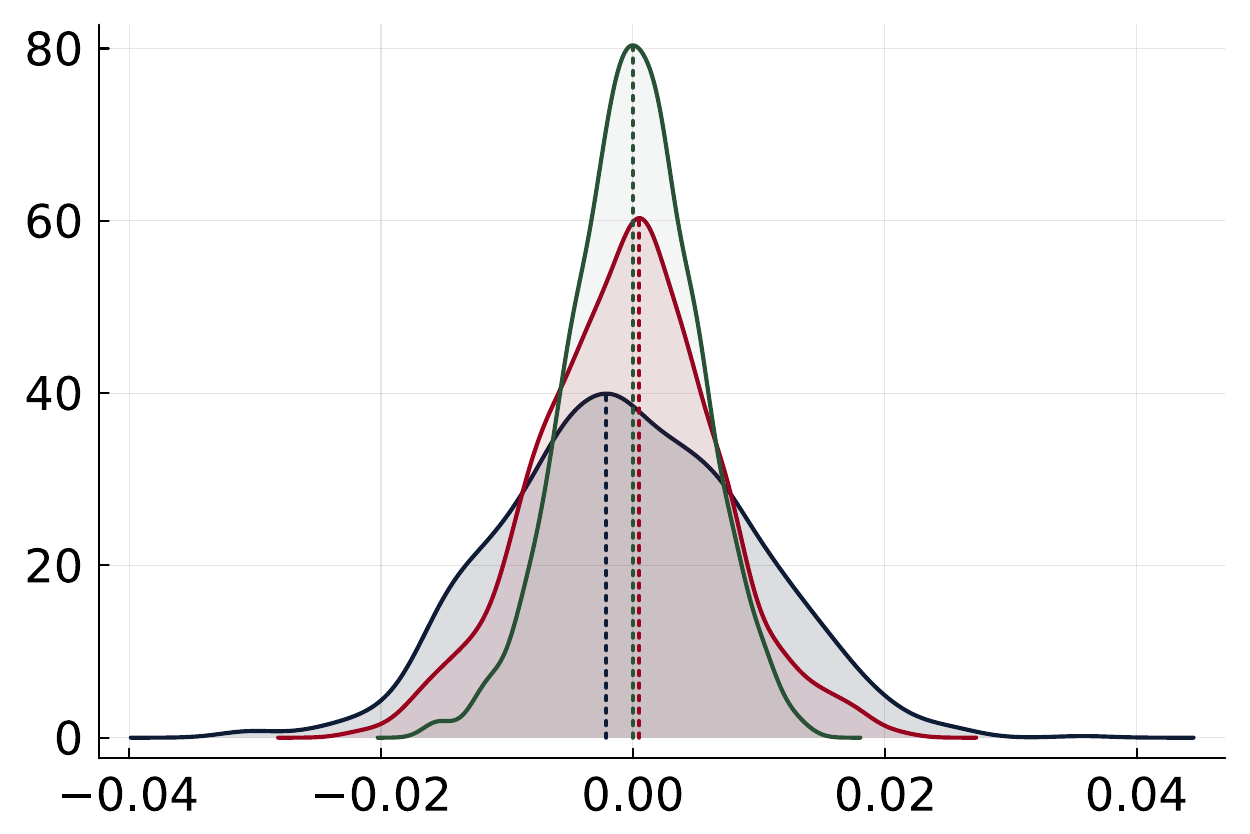}
    \caption{$\hat \gamma_4$}
\end{subfigure}
\hfill
\begin{subfigure}{0.32\textwidth}
    \centering
    \includegraphics[width = \textwidth]{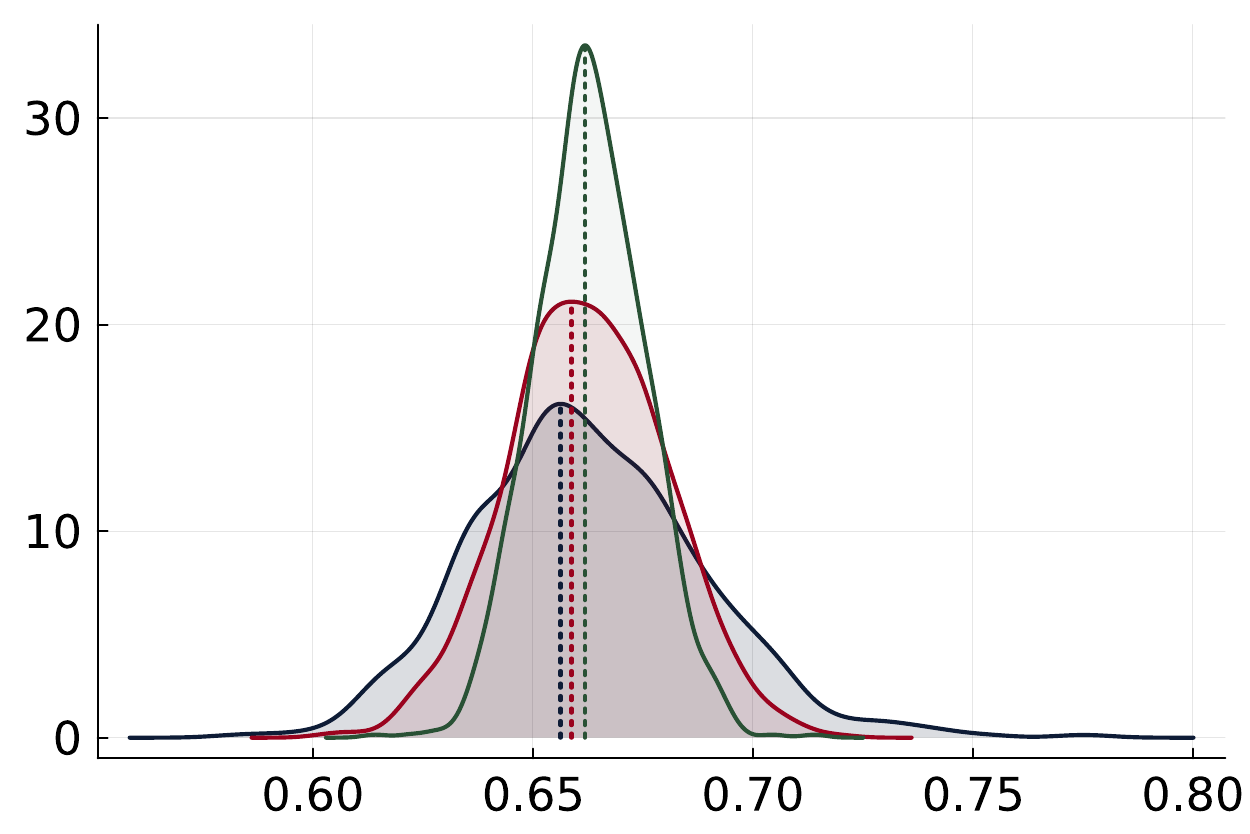}
    \caption{$\hat\delta$}
\end{subfigure}

    \caption{Top row: imputation estimates from interval censored exponential Hawkes process with parameter $(\nu, \eta, \beta) = (2.0, 0.6, 2.0)$ for $T \in\{500,\,1,\!000\}$. Bottom row: Select $\mathrm{NBAR}(10)$ estimates and dispersion estimate $\hat\phi$ for interval censored Hawkes process with $\mathrm{Gamma}(\alpha,\beta)$ kernel and parameter $(\nu, \eta,\alpha, \beta) = (2.0, 0.6, 1.5,0.25)$ for $T \in\{ 500,\,1,\!000\}$. Censoring intervals are of width $\Delta = 1.0$ in both cases.}
    \label{fig:convergence}
\end{figure}

Figure~\ref{fig:convergence} displays imputation and $\mathrm{NBAR}(10)$ estimates of $M = 500$ simulated sample paths at different censoring times, $T$, for $\Delta = 1.0$. The estimates appear to converge to a limit and are approximately normally distributed, which is expected for MLEs in sufficiently regular models. As expected, the imputation estimates do not converge to the true parameter. 

Despite the apparent convergence of the $\mathrm{NBAR}(p)$ estimates, for fixed $p$, the estimates cannot be asymptotically sufficient for offspring densities with unbounded support. Taking $p\to\infty$ does not necessarily guarantee asymptotic sufficiency, as the $\mathrm{NBAR}$ model may not be rich enough to capture higher order statistical properties of the discretely observed Hawkes process. The empirical performance of the NN estimator serves as evidence that the $\mathrm{NBAR}$ estimates are highly informative.

\subsection{Numerical Evidence of Consistency of the NN estimator}
For fixed $\Delta$, the NN estimator behaves as though it converges in mean-squared error (MSE) at the rate $\cc O(\frac{1}{K})$. Table~\ref{tab:sqrt_T} shows the NN estimates of simulated discretely observed Hawkes process data for $\Delta = 0.1$, for different censoring times, $T$ (thus $K = 10T$). The bias and standard error of the estimates decrease with $K$, with the rate of decrease in MSE congruent with the $\mathcal{O}(\frac{1}{K})$ behaviour of MSE with $\sqrt{K}$-consistent estimators. The convergence of the summary statistics appears to yield convergence of the NN estimators, as desired.

\begin{table}
    \centering
    \caption{NN estimates with exponential kernel, $\Delta = 0.1$, and varying $T$. NN model: two hidden layers with 64 and 32 nodes, respectively.}
    \setlength{\tabcolsep}{2.75pt}
    \begin{tabular}{lccccc}\toprule
     & & $\nu$ &$\eta$ &$\beta$ & \multirow{2}{*}{$\mathrm{MSE}$}  \\
     \cmidrule(lr){3-5}
     & & $2.0$ & $0.6$ & $2.0$ \\ 
    \midrule
\multirow{3}{*}{$T\, =\, 250$} & Est & 2.167 & 0.568 & 2.287 & \multirow{3}{*}{$0.260$} \\
& SE  & 0.371 & 0.077 & 0.726 \\
& CP & 0.966 & 0.950 & 0.938 \\
\midrule
\multirow{3}{*}{$T\, =\, 500$}
& Est & 2.098 & 0.583 & 2.154 & \multirow{3}{*}{$0.114$} \\
& SE  & 0.309 & 0.057 & 0.458 \\
& CP & 0.952 & 0.968 & 0.948 \\
\midrule
\multirow{3}{*}{$T\, =\, 1,\!000$}
& Est & 2.046 & 0.591 & 2.067 & \multirow{3}{*}{$0.047$} \\
& SE  & 0.214 & 0.046 & 0.308  \\
& CP & 0.958 & 0.958 & 0.934 \\
    \bottomrule
\end{tabular}
    
    \label{tab:sqrt_T}
\end{table}

\section{Amortisation relative to the length of the observed sample path}
Consider a NN $F_1$ that is trained on data simulated to censoring time $T_1$, with censoring intervals of fixed width $\Delta > 0$, yielding $K_1 = \lfloor T_1/\Delta\rfloor$ observations. For sample paths $n_{1:K_1}$ generated from parameter $\tht$, the summary statistics $\bs s(n_{1:K_1})$ will be distributed around their mean, $\bb E_\tht [\bs s(n_{1:K_1})]$. Since we observe numerical convergence of the imputation and NBAR estimates (Figure~\ref{fig:convergence}), for sufficiently large $K_1$, we have $\bb E_\tht[\bs s(n_{1:K_1})] \approx C_\tht$ for some limiting vector $C_\tht$. 

Suppose then one has a sample path $n_{1:K_2}$ generated from $\tht_0$, with $K_2 \neq K_1$. For large $K_2$, the summary statistic $\bs s(n_{1:K_2})$ will similarly be centred around $\bb E_{\tht_0}[\bs s(n_{1:K_2})]\approx C_{\tht_0}$. The trained NN $F_1$ accurately estimates $\tht_0$ from $\bs s(n_{1:K_2})$. Thus, the amortisation of point estimation by the NN relative to the number of observations is justified by the convergent behaviour of the summary statistics.

The quantile estimation procedure is amortised relative to the number of observations under some additional assumptions. Let $\tht_{\bs s}$ denote the posterior distribution $\tht \mid \bs s(n_{1:K})$. The classical Bernstein-von Mises theorem states that under suitable regularity conditions, for the true parameter $\tht_0$, the posterior distribution $\sqrt{K}(\tht_{\bs s} - \tht_0)$ converges to a Gaussian in total variation. See Chapter 10 of \cite{vaartAsymptoticStatistics1998} for detailed conditions of the theorem. We make the following assumption.
\begin{assumption}
    The conditions of the Bernstein-von Mises theorem are satisfied by the posterior $\tht_{\bs s}$.
\end{assumption}
The conditions are challenging to verify directly due to the intractable form of $p_\tht(n_{1:K})$. However, the observed $\sqrt{K}$-consistency of the NN estimation procedure (Table~\ref{tab:sqrt_T}) is indicative that the rate of convergence suggested by the Bernstein-von Mises theorem holds for our NN estimator. Letting $\hat q_{\tau, K_1}$ be the $\tau$-quantile estimate produced by the trained model $F_1$, we make the following adjustment:
\begin{align*}
    \hat{q}_{\tau, K_2} \ &= \ \hth \ + \ \sqrt{\frac{K_1}{K_2}}\big(\hat{q}_{\tau,K_1} \ - \ \hth \big).
\end{align*}
Intuitively, we are modulating the gap between the estimated median of the posterior and the estimated quantile by the theoretical factor $\sqrt{K_1/K_2}$. Table~\ref{tab:amortisation} presents the outcome of this approximation. A NN is trained on data with $T = 1,\!000$ and $\Delta = 0.1$. Data simulated from paths for which $T = 500$ and $T = 2,\!000$ are estimated using the trained model. For comparison, the simulated paths are also estimated using NNs trained on data simulated to the true censoring time. The parameter estimates are similar between the two NNs, respectively, and the quantile adjustment procedure yields well calibrated credible intervals.

\begin{table}[ht]
    \centering
    \caption{Top row: NN estimates from NNs trained on the true censoring time, $T = 500,\, 2,\!000$, respectively. Bottom row: NN estimates from NNs trained on data with censoring time $T = 1,\!000$, using the adjustment procedure. Censoring interval width $\Delta = 0.1$, all NNs having two hidden layers with 64 and 32 nodes, respectively.}
    \begin{tabular}{lcccc|cccc}\toprule
    && \multicolumn{3}{c|}{$T = 500$} && \multicolumn{3}{c}{$T = 2,\!000$}\\ 
    \midrule
 && $\nu$  &$\eta$ & $\beta$  && $\nu$  &$\eta$ & $\beta$\\
     \midrule
    && $2.0$ &$0.6$ & $2.0$ && $2.0$ &$0.6$ & $2.0$ \\ 
    \midrule
    \multirow{3}{*}{NN trained using true $T$} & Est & 2.081 & 0.581 & 2.045 & Est & 2.007 & 0.597 & 2.009 \\
& SE  & 0.277 & 0.056 & 0.422 & SE  & 0.164 & 0.035 & 0.211 \\
& CP & 0.952 & 0.962 & 0.946 & CP & 0.940 & 0.930 & 0.958 \\
    \midrule
    \multirow{3}{*}{NN trained using $T=1,\!000$} & Est & 2.094 & 0.581 & 2.031 & Est & 2.019 & 0.595 & 2.050\\
    & SE  & 0.294 & 0.060 & 0.442 & SE  & 0.165 & 0.032 & 0.209 \\
    & CP & 0.946 & 0.952 & 0.954 & CP & 0.930 & 0.930 & 0.948 \\
     
    \bottomrule
\end{tabular}
    \label{tab:amortisation}
\end{table}

\end{document}